\pdfoutput=1
\PassOptionsToPackage{hyperref}{%
  pdfauthor={Marc Hennes and Julien Tailleur and Gaëlle Charron and Adrian Daerr},
  pdftitle={Active depinning of bacterial droplets: the collective surfing of Bacillus subtilis},
  pdfsubject={Physics,biophysics,soft matter},
  pdfkeywords={Wetting, Bacterial motility, Collective motion}}

\documentclass[9pt,twocolumn,twoside]{pnas-new}

\templatetype{pnasresearcharticle} 

\newcounter{publishedpage}
\makeatletter
\fancypagestyle{firststyle}{
   \fancyfoot[R]{\footerfont PNAS\hspace{7pt}|\hspace{7pt}\textbf{June 6, 2017}\hspace{7pt}|\hspace{7pt}vol. 114\hspace{7pt}|\hspace{7pt}no. 23\hspace{7pt}|\hspace{7pt}\textbf{{\setcounter{publishedpage}{\value{page}}\addtocounter{publishedpage}{5957}\thepublishedpage}\textendash 5963}}
   \fancyfoot[L]{\footerfont\@ifundefined{@doi}{}{\@doi}}
}
\makeatother
\setboolean{displaywatermark}{false}

\graphicspath{{figures/},{supplementary/}}
\usepackage[retain-unity-mantissa = false,exponent-product = \cdot]{siunitx}
\usepackage[version=3]{mhchem}
\usepackage{wrapfig}

\DeclareUrlCommand\email{}
\newcommand{\Bsubtilis}{\textit{B. subtilis}}
\newcommand{\surfactin}{\textit{surfactin}}

\newcommand{\Bond}{\text{Bo}}
\newcommand{\OD}{\text{OD}}
\newcommand{\water}{\text{\ce{H2O}}}
\newcommand{\bact}{\text{bact.}}
\newcommand{\Capillary}{\text{Ca}}

\newcommand{\SUPPwatercontamination}{SF1}
\newcommand{\SUPPwatercontaminationSR}{SF2}
\newcommand{\SUPPcontactangles}{SF3}
\newcommand{\SUPPdryingtime}{SF4}
\newcommand{\SUPPglucSDS}{SF5}
\newcommand{\SUPPopticaldensity}{SF6}
\newcommand{\SUPPbactdrop}{SF7}
\newcommand{\SUPPhplccomparison}{SF8}
\newcommand{\SUPPhplccalibration}{SF9}
\newcommand{\SUPPpetridishsliding}{SF10}
\newcommand{\SUPPhplcconcentrations}{ST1}
\newcommand{\SUPPmovieSurf}{SM1}
\newcommand{\SUPPmovieInducedSN}{SM2}
\newcommand{\SUPPmovieInducedWater}{SM3}
\newcommand{\SUPPmovieInducedSR}{SM4}
\newcommand{\SUPPmovieSurfactinSlide}{SM5}
\newcommand{\surfactantstrain}{SSB\,1071}

\title{Active depinning of bacterial droplets:
       the collective surfing of Bacillus subtilis}

\author[a]{Marc Hennes}
\author[a]{Julien Tailleur}
\author[a]{Ga{\"e}lle Charron}
\author[a,1]{Adrian Daerr}

\affil[a]{Laboratoire Mati{\`e}re et
  Syst{\`e}mes Complexes (MSC) UMR CNRS 7057, University Paris
  Diderot, 75205 Paris cedex 13, France}

\leadauthor{Hennes} 

\significancestatement{How bacteria propagate and colonise new environments is
  of paramount importance e.g. to food processing, transport and
  storage. In this article we show that \textit{Bacillus subtilis} can
  turn millimetric water droplets into vehicles to quickly cover large
  distances. To this end they overcome the enormous capillary forces
  that prevent the sliding of water droplets even on vertical surfaces
  such as refrigerator walls. \Bsubtilis{} does so by actively
  unpinning the surrounding liquid to cause the collective surfing of
  the entire community on slopes as small as \ang{0.1}. We
    show this depinning to rely on a variety of mechanisms, that could
    each be relevant to the motility of microorganisms in partially
    humid environments such as unsaturated soils.}

\correspondingauthor{\textsuperscript{1}To whom correspondence should
  be addressed. E-mail: \email{adrian.daerr@univ-paris-diderot.fr}\smallskip\\
This article contains supporting information online at \url{www.pnas.org/lookup/suppl/doi:10.1073/pnas.1703997114/-/DCSupplemental}.}
\authorcontributions{MH, JT, GC, and AD designed research;
  MH and AD performed research; 
  MH, JT, GC, and AD analysed data; 
  MH, JT, and AD wrote the paper.}
\authordeclaration{The authors declare no conflict of interest.}


\begin{abstract}
  How systems are endowed with migration capacity is a fascinating
  question with implications ranging from the design of novel active
  systems to the control of microbial populations.
  Bacteria, which can be found in a variety of environments, have
  developed among the richest set of locomotion mechanisms both at the
  microscopic and collective levels.
  Here, we uncover experimentally a new mode of collective bacterial
  motility in humid environment through the depinning of bacterial
  droplets.
  While capillary forces are notoriously enormous at the bacterial
  scale, even capable of pinning water droplets of millimetric size on
  inclined surfaces, we show that bacteria are able to harness a
  variety of mechanisms to unpin contact lines, hence inducing a
  collective slipping of the colony across the surface.
  Contrary to flagella-dependent migration modes like swarming we show
  that this much faster `colony surfing' still occurs in mutant
  strains of \textit{Bacillus subtilis} lacking flagella.
  The active unpinning seen in our experiments relies on a
  variety of microscopic mechanisms which could each play an
  important role in the migration of microorganisms in humid
  environment.
\end{abstract}

\dates{This is the preprint as submitted to PNAS. For the final published version see the PNAS web site at \url{www.pnas.org/cgi/doi/10.1073/pnas.1703997114}}
\doi{\url{www.pnas.org/cgi/doi/10.1073/pnas.1703997114}}

\begin{document}


\maketitle
\thispagestyle{firststyle}
\ifthenelse{\boolean{shortarticle}}{\ifthenelse{\boolean{singlecolumn}}{\abscontentformatted}{\abscontent}}{}

\providecommand{\body}{}
\body

\dropcap{C}ollective behaviours in bacterial colonies, from swarming
and bacterial turbulence~\cite{wensink2012meso} to
biofilm~\cite{lopez2010biofilm,bridier2011biofilm} and pattern
formation~\cite{budrene1991complex, brenner1998physical,
  cates2010arrested, liu2011sequential}, are among the most
fascinating problems at the crossroads of biology, chemistry and
physics. They raise fundamental questions, such as the emergence of
complex behaviours from simple microscopic dynamics, and are of
practical importance due to their medical
impact~\cite{hall2004bacterial}. Thanks to studies carried out in
simplified, controlled environments, much is known about the swimming
of bacteria in three dimensional fluids~\cite{berg2008coli}. A lot of
effort has thus been devoted over the past few years to the study of
bacterial motion in more complex environments, in particular near
interfaces~\cite{drescher2011fluid,tuval2005bacterial,zhang2010upper}.

A prototypical case study is the spreading of \textit{Bacillus
  subtilis} colonies on the surface of agar
gels~\cite{fujikawa1989fractal,ben1994generic}. By changing the
nutrient and agar concentrations, very diverse colony morphologies are
observed, ranging from simple diffusive spreading to the formation of
fractal dendritic structures. The motion of bacteria on hard gel
substrates vividly differs from the unconstrained swimming in liquid:
isolated bacteria indeed cannot move on agar gels since their
propulsion forces are much weaker than the capillary forces they
experience from the surrounding liquid film. Colony migration remains
possible but relies on cooperativity to overcome pinning, as in the
dense mass swarming that many bacteria
exhibit~\cite{harshey2003,julkowska2004}.

The importance of capillary forces at small scales is familiar to us
all from the observation of glass windows after the rain, or the walls
of a fridge or cold storage facility, on which small drops remain
stuck. A drop can indeed only slide when the driving gravitational
force overcomes the capillary pinning force. Similarly, small water
drops of, say, \SI{2}{\micro\litre} remain stuck at the surface of an
agar gel, at any tilt angle. Here we show that bacteria are able to
unpin such droplets, leading in practice to the collective `surfing'
of the entire colony. Surprisingly, this happens on slopes \textit{as
  small as \ang{0.1}}, commonly encountered on any generic surface,
and leads to migration speeds well above that of mass swarming.

\begin{figure}%
\includegraphics[width=0.99\columnwidth]{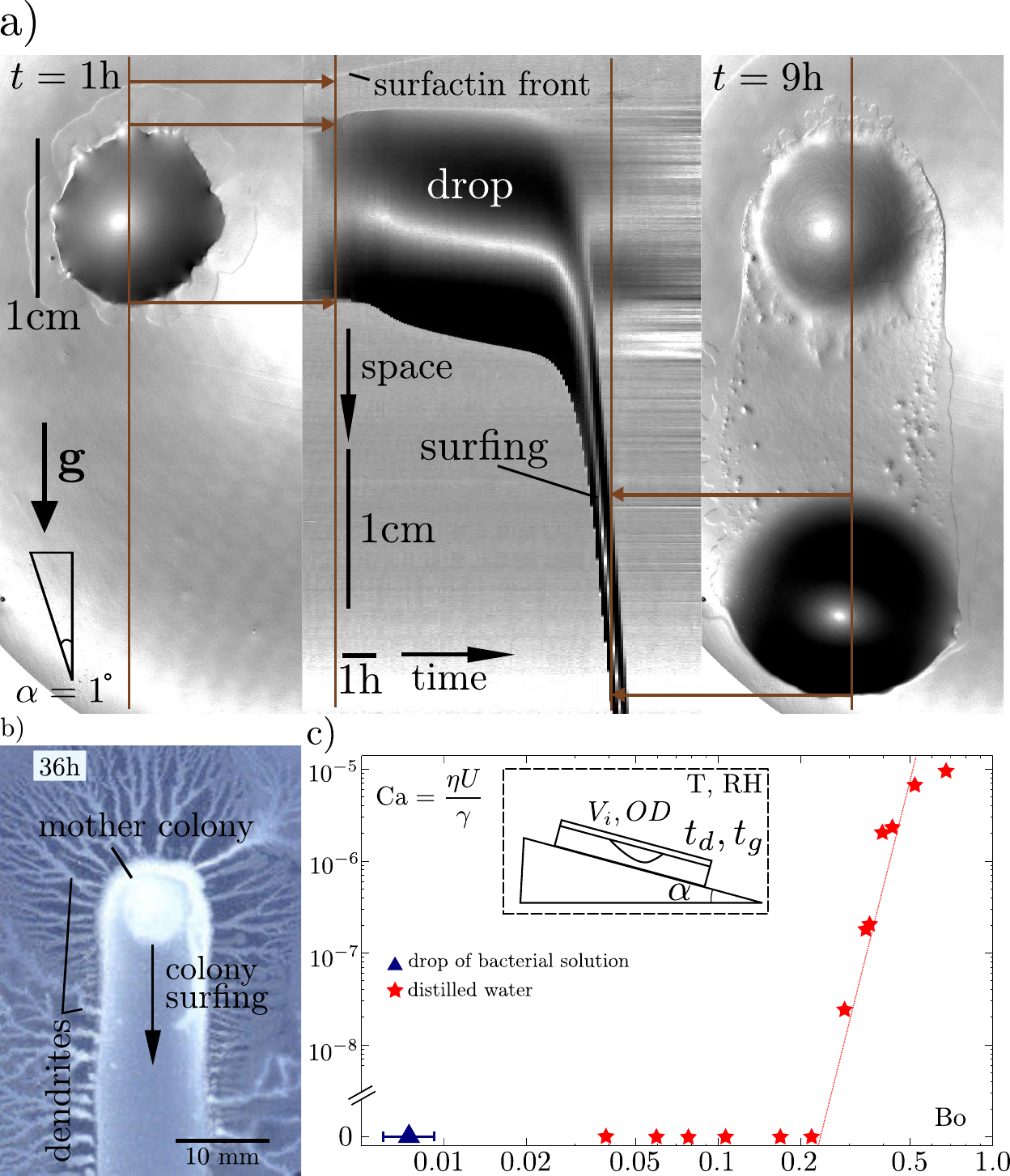}
\caption{\textbf{a)} Top view of colony surfing where grey levels
  indicate the local slope. A $\SI{2}{\micro\litre}$ drop is still
  immobile at $t=\SI{1}{hour}$ after its deposition (left). The onset
  of slipping motion at $t=\SI{7}{hour}$ is visible on the central
  kymograph which represents the time evolution of a cut through the
  drop along the direction of motion (brown line). The gel remains
  swollen around the point of deposition (upper dark region, right).
  \textbf{b)} 36h after deposition, the bacteria which have been
  carried by the sliding drop have invaded the whole drop trajectory
  and its surroundings. \textbf{c)} Dimensionless velocity vs
  gravitational pull of water drops, measured by their Capillary and
  Bond numbers (red stars). The initial Bond number of the
  bacteria-laden drop shown in (a) (\num{0.003}, blue triangle) is
  orders of magnitude below the critical Bond number
  $\Bond_{\rm c}^{\water} \approx \num{0.25}$ at which water drops
  start sliding. }%
\label{fig:kinetics+Bond}
\end{figure}

\section*{Results}

\subsection*{Active depinning of bacterial droplets} A
\SI{2}{\micro\litre} drop of a suspension of \Bsubtilis{} grown to an
optical density $\OD=0.27$ is deposited on top of a \SI{0.7}{\percent}
agar gel substrate with nutrients, and incubated in a climate chamber
at fixed temperature and relative humidity (T=\SI{30}{\celsius},
RH=\SI{70}{\percent}, see \textit{Materials and Methods}).
Fig.~\ref{fig:kinetics+Bond}a and movie~\SUPPmovieSurf{} show the
typical evolution of the drop for a gel surface inclination of
$\alpha = \ang{1}$. During the first hour, it exhibits a slight
isotropic spreading, typical of aqueous surfactant solutions on
hydrogels~\cite{lee2008,banaha2009}. This is expected since our
\Bsubtilis{} strain is known to produce \surfactin{}, a cyclic
lipopeptide which strongly reduces surface tension~\cite{arima1968,
  peypoux1999}. The surfactant also causes some swelling of the gel as
it spreads outwards, producing a visible `surfactant
ring'~\cite{debois2008} that grows radially. Around seven hours after
deposition, however, the drop starts sliding in the direction of
inclination. Bacteria move along with the drop as it slides, hence the
name \textit{colony surfing}. The drop velocity rapidly increases to
reach \SI{1.5}{\centi\metre\per\hour}, almost an order of magnitude
larger than the speed of swarming of our \Bsubtilis{} strain on flat
gels. This surfing motion persists at roughly constant drop diameter
and velocity until the drop reaches the rim of the Petri dish.
Interestingly, bacteria continuously exit from the drop as it slides
so that they rapidly colonise the whole drop trajectory and its
surrounding (Fig.~\ref{fig:kinetics+Bond}b).

\begin{figure}
  \centering
  \includegraphics[width=.6\columnwidth]{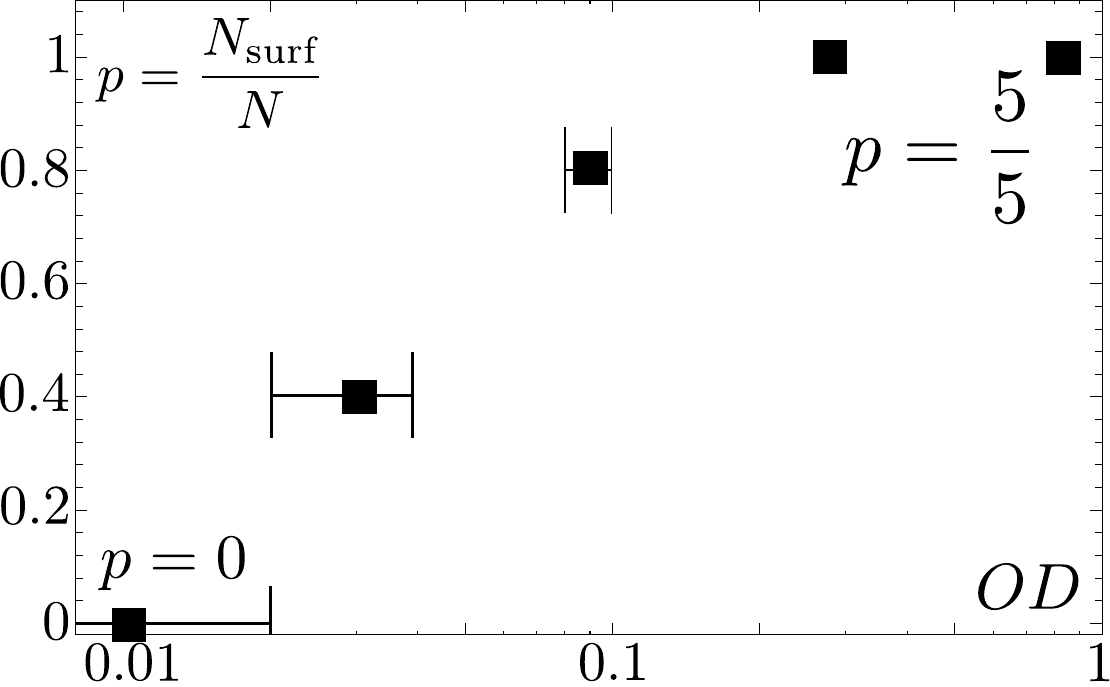}
  \caption{Fraction of colony surfing events for different bacterial
    concentrations in the deposited drop. Out of five experiments,
    none show surfing for initial optical densities below
    $\OD = \num{0.02}$, while surfing always occurs at
    $\OD = \num{0.27}$ and above.}
  \label{fig:varyOD}
\end{figure}

The mobility of the bacteria-laden drops is striking when compared to
droplets of water or pure nutrient solution. Water droplets indeed
only start sliding on slopes \textit{two orders of magnitude larger}
than bacterial droplets of the same volume. This is best quantified
using the ratio of gravitational to capillary forces, using the
so-called Bond number
\begin{equation}\label{eq:mynameisJames}
  \Bond 
  = \frac{\rho V  g \sin \alpha}{V^{1/3}\gamma}
\end{equation}
where $\rho V$ and $V^{1/3}$ are the drop mass and typical width,
$g \sin\alpha$ the effective gravity, and $\gamma$ the surface
tension. We always observe the sliding of bacteria-laden droplets with
an initial Bond number of $\Bond\simeq \num{3e-3}$ whereas water drops
only start sliding for Bond numbers larger than
$\Bond_{\rm c}^{\water} \approx 0.25$ (Fig.~\ref{fig:kinetics+Bond}c).
The transition from an immobile water drop to a sliding bacteria-laden
one occurs continuously as the initial concentration of bacteria is
varied (Fig.~\ref{fig:varyOD}). We never observe colony surfing for
bacterial densities below a threshold of $\OD \approx 0.02$ within the
first 12 hours of recording. The \surfactin{} ring is still
observable, but its formation is delayed by several hours. For higher
concentrations, the probability of colony surfing increases, and was
observed in five out of five runs for $\OD$s above $\approx 0.2$.

\begin{figure}
\centerline{\includegraphics[width=0.9\columnwidth]{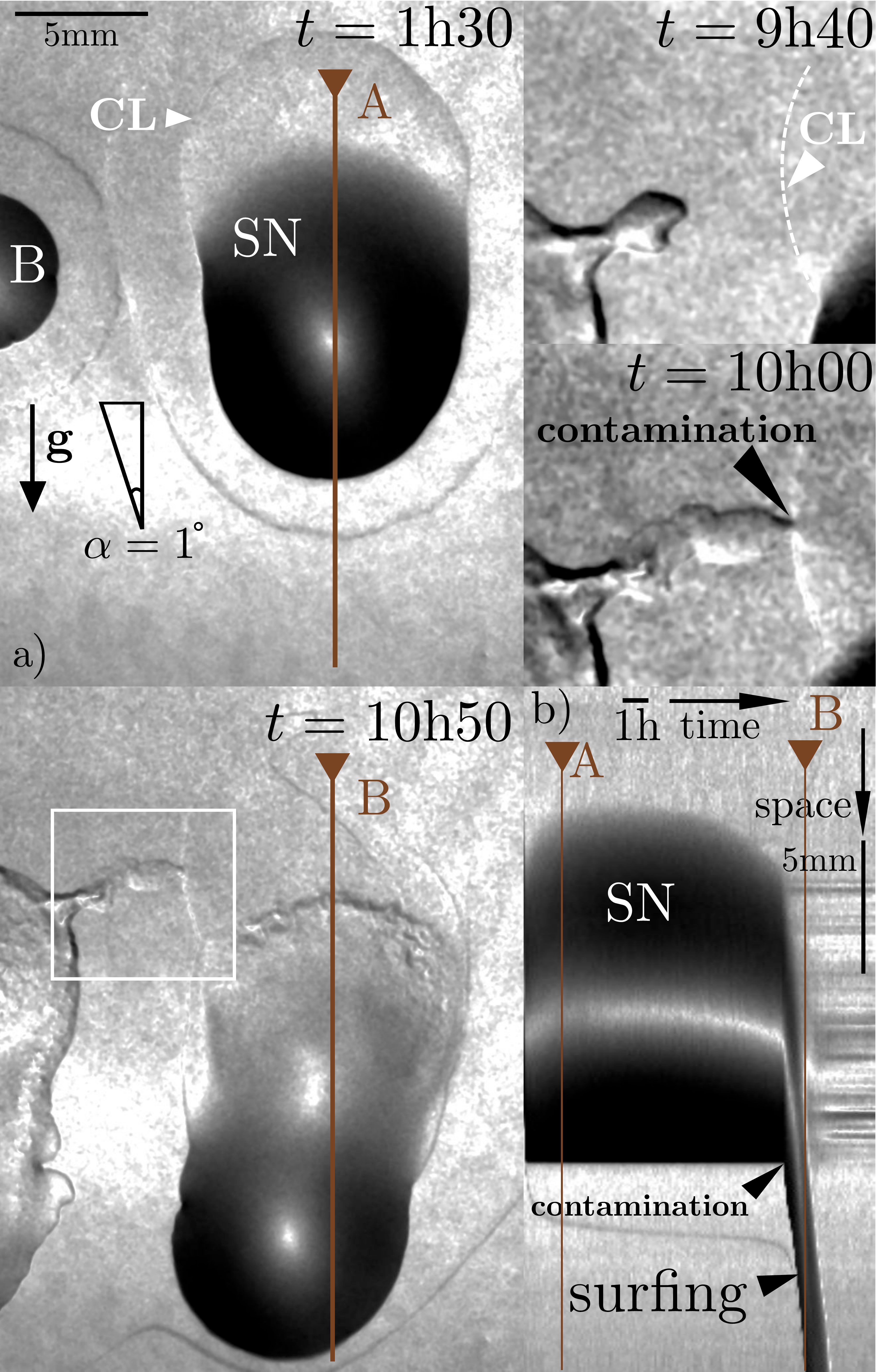}}
\caption{\textbf{a)} Snapshots of induced colony surfing, where the
  grey-level indicates the local slope. Two drops are deposited side
  by side. From the right drop the bacteria were removed by
  centrifugation and filtration. The bacteria-laden drop (B,
  \SI{1}{\micro\litre}) starts sliding after 8 hours while the drop
  containing only supernatant (SN, \SI{10}{\micro\litre}) remains
  immobile despite its much larger volume. The magnified details at
  the top right (white rectangle) show how at
  $t \approx \SI{10}{\hour}$, a dendrite of swarming bacteria
  contaminates the immobile drop (white arrow marks contact line CL),
  which starts sliding almost instantaneously. \textbf{b)} Kymograph
  of a longitudinal cut through the SN drop showing the very sudden
  onset of motion of the drop of supernatant after its contamination.
  See suppl. movie~\SUPPmovieInducedSN.}%
\label{fig:inducedsurf}
\end{figure}

The bacterial medium does not solely contain bacteria but also
nutrients and bacterial products including in particular \surfactin{}
and exopolysaccharides. To test whether the colony surfing is really
due to bacteria we remove them from an $\OD = 0.27$ suspension by
centrifugation and filtration, and start experiments where two drops,
with and without bacteria, are deposited side by side
(Fig.~\ref{fig:inducedsurf} and supplementary
movie~\SUPPmovieInducedSN). While the drop containing bacteria starts
sliding after \SI{8}{\hour}, the bacteria-free drop initially remains
immobile. After \SI{10}{\hour}, a small dendrite of bacteria
contaminates the second drop, which starts sliding almost immediately
(the time resolution is given by the \SI{5}{\minute} interval at which
images are taken). This highlights the role played by bacteria in our
set-up. Note that contamination can also induce slipping of pure water
droplets (Supplementary Fig.~\SUPPwatercontamination{} and
movie~\SUPPmovieInducedWater).

\begin{figure}
  \centering
  \includegraphics[width=.6\columnwidth]{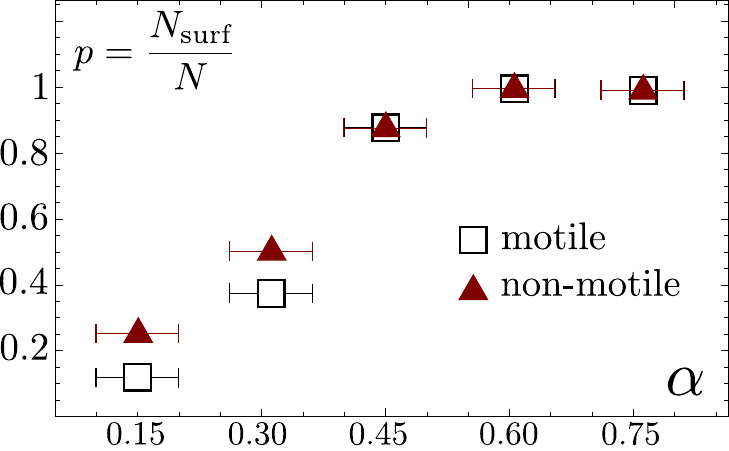}
  \caption{Fraction of colony surfing events $p = N_{\text{surf}}/N$,
    out of $N=8$ runs for each data point, for $\SI{2}{\micro\litre}$
    drops (initial $\OD=0.27$) of a motile (\surfactantstrain) and
    non-motile (OMG\,954) strain at different gel inclinations. The
    two strains show no statistically significant difference. The
    uncertainty of the tilt angle measurement is estimated to
    $\Delta \alpha = \pm \ang{0.05}$.}%
  \label{fig:motility}
\end{figure}

\subsection*{The role of bacterial propelling forces}
The depinning of the contact line (drop perimeter) arises when driving
forces exceed the capillary forces. A natural explanation of the
depinning of the contact line could thus be an extra
  contribution to the driving force due to bacterial
  motion~\cite{nikola2016}: the self-propelling forces combined with
  gravity could, together, overcome the pinning force. Active forces
  are also known to accelerate the relaxation of polymer
  networks~\cite{humphrey2002}. This could affect the bulk rheology of
  the droplet itself, modify the dynamics of gel polymer desorption
  near the contact line or accelerate the visco-plastic stress
  relaxation in its vicinity, hence affecting the contact line
  mobility~\cite{kajiya2013stickslip,cohenstuart2006}. Finally, active swimming
  strongly enhances transport and mixing within the droplet, notably
  of nutrients and oxygen~\cite{sokolov2009}, which could facilitate
  the collective surfing of motile colonies in comparison to
  non-motile ones. In order to determine whether bacterial motility
  has an impact on colony sliding, we compared the surfing likelihood
of a motile and a non-motile strain. The strains differ only in the
deletion of the \textit{hag} gene coding for \textit{flagellin}, the
protein sub-unit required for the assembly of the flagella. As shown
in Fig.~\ref{fig:motility}, the fraction of surfing events is similar
for both strains, showing that the collective colony surfing does not
require individual motility. The sole alternative are thus a decrease
of capillary forces, or an increase of the gravitational pull.

\begin{figure}
  \centering
  \includegraphics[width=.7\columnwidth]{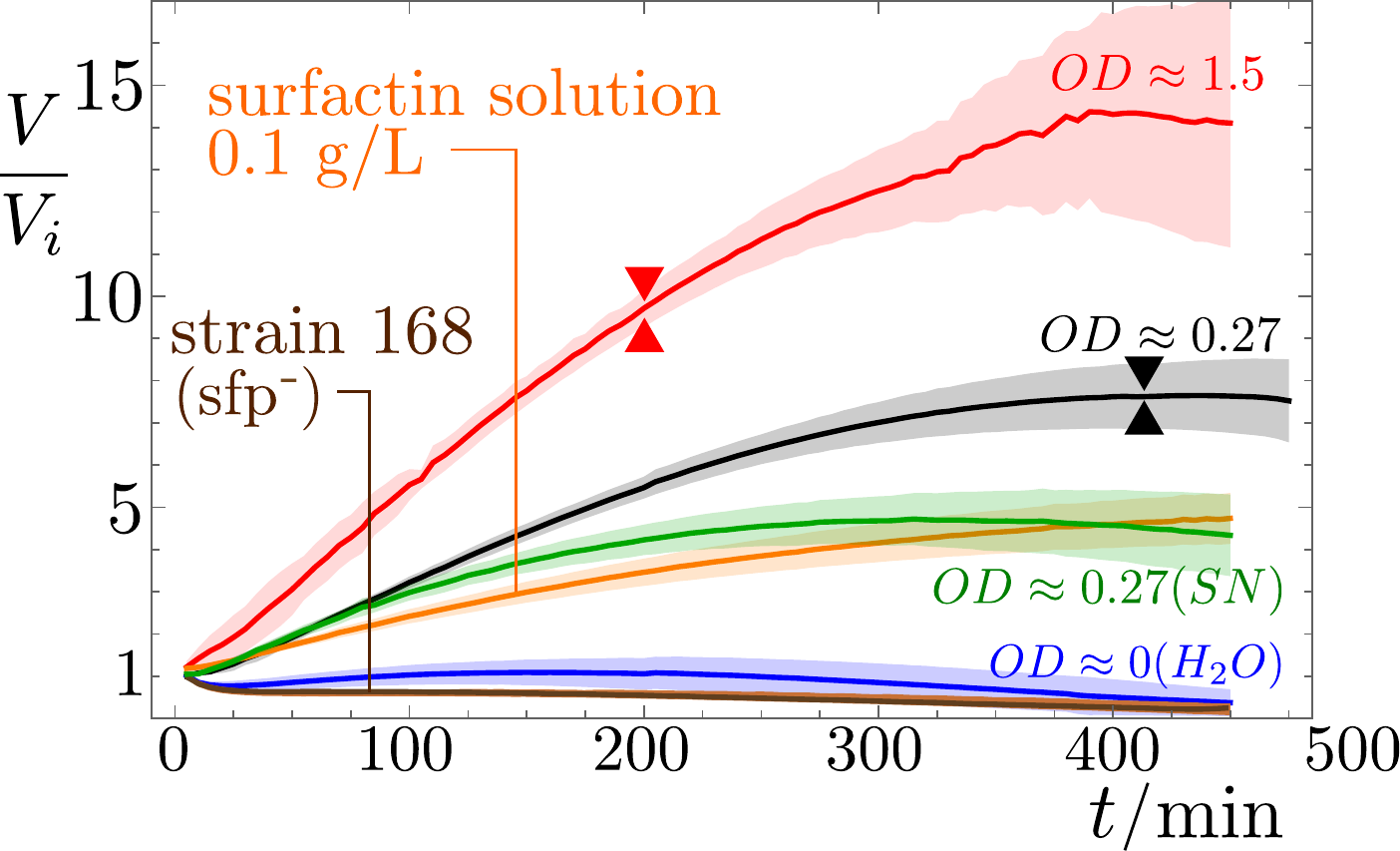}
  \caption{Relative volume change of deposited drops of bacteria,
    supernatant (SN), solutions of commercial surfactin, and pure
    water, with respect to their deposited volume
    $V_i=\SI{2}{\micro\litre}$. Higher bacterial optical densities at
    deposition result in stronger volume increase, likely because the
    surfactant production rate is higher. Removing the bacteria
    through centrifugation from an $\OD=0.27$ suspension and
    filtration just before deposition (supernatant, SN) does not
    affect the initial volume growth rate, but the subsequent
    evolution differs from the bacterial suspension (green vs. black
    curves). Pure \surfactin{}-water drops also cause a volume
    increase, while no volume increase is measured for drops of pure
    water, nor for the non-surfactin-producing strain 168. The onset
    of surfing \textemdash{} if it occurs \textemdash{} is marked by
    double triangles. Solid lines stem from averaging over several
    experiments; the corresponding standard deviation is indicated by
    shaded areas.}\label{fig:volume}
\end{figure}

\subsection*{Volume increase of bacterial droplet}
As surprising as it may sound, bacteria are indeed able to increase
the gravitational pull exerted on the surrounding droplet, by pumping
water out of the environment. We show in Fig.~\ref{fig:volume} the
evolution of the drop volume, measured by means of a Moir\'e fringe
deformation method (see \textit{Materials and Methods}). Drops with a
high density of bacteria show a strong volume increase prior to colony
surfing. For the \SI{2}{\micro\litre} drop of $\OD \approx 0.27$
described above, the volume grows monotonically for the first five to
six hours to reach around eight times its initial value, at which
point the drop starts sliding. For a higher initial concentration,
$\OD \approx 1.5$, the drop shows a quicker volume increase, reaching
a volume fifteen times its initial value and starts sliding much
earlier (\SI{3.5}{\hour} after deposition).

The volume increase of pure water drops without bacteria in the same
conditions is nil or negligible. On the other hand, droplets of an
$\OD = 0.27$ suspension from which bacteria are removed through
centrifugation and filtration still show a significant volume
increase. At this stage the supernatant (SN) solution already contains
large amounts of \surfactin{} (Suppl. Fig.~\SUPPhplccalibration{} and
Table~\SUPPhplcconcentrations), along with other products including
exopolysaccharides and, of course, nutrients. A possible explanation
is that the concentration gradients of chemicals produced by bacteria
generate inward osmotic flows. This hypothesis is validated by the
fact that drops of aqueous solution containing either pure
\surfactin{}, or the anionic surfactant sodium dodecyl sulfate (SDS,
Fig.~\SUPPglucSDS), grow to four-to-eight times their initial volume
within a four-to-seven-hour period. We believe \surfactin{} to be the
main responsible of these osmotic flows since the strain 168 lacking
\surfactin{} synthesis, because of a frameshift mutation in the
\textit{sfp} gene, is never seen to produce colony surfing at any
$\OD$, and the corresponding deposited drops do not significantly
inflate (brown line in Fig.~\ref{fig:volume}). This absence of volume
increase also highlights the negligible role of cell division in this
process. The density of bacteria indeed remains
  very low throughout the experiment, as we confirmed by direct
  measurement (Fig.~\SUPPbactdrop). Note however, that this osmotic
growth is not limited to surfactants: a drop containing simply
\SI{50}{\percent} glucose also undergoes a similar volume increase
(Fig.~\SUPPglucSDS). This shows that this active pumping of water from
the environment could be seen beyond the sole case of
surfactant-producing bacteria reported here.

Finally, note that evaporation limits the time window for colony
surfing in our setup. Indeed, under the \SI{70}{\percent} relative
humidity in the climate chamber commonly used in swarming experiments,
the gel slowly evaporates at a rate of about $\epsilon_{evap} \simeq$
\SI{0.3}{\micro\litre\per\hour\per\square\centi\metre}, as measured by
weighing over a duration of \SI{72}{\hour}. As time goes on,
concentration gradients decrease as the surfactin molecules diffuse in
the gel so that the inward osmotic flow decreases. Evaporation will
eventually dominate, so that a drop only reaches a high enough Bond
number and slides if its volume increase occurs within the first few
hours of the experiment. Similarly, letting the gel dry in open air
before the experiment makes it more difficult for bacteria to extract
water out of the gel, hence limiting the overall volume increase of
deposited drops (Fig.~\SUPPdryingtime{}). This probably explains why
the active unpinning of bacterial droplets has never been reported
before. On the contrary, no time limit should apply to colony surfing
in saturated atmosphere, e.g. in the presence of a nearby water
reservoir or in partially soaked porous media.

The volume increase of the droplet generates a proportional increase
of gravitational pull, while capillary pinning forces increase only
with the contact line length, which scales as $V^{1/3}$. The overall
effect of the volume increase is thus to facilitate colony surfing by
increasing the Bond number.

\begin{figure}
\centering
\includegraphics[width=.7\columnwidth]{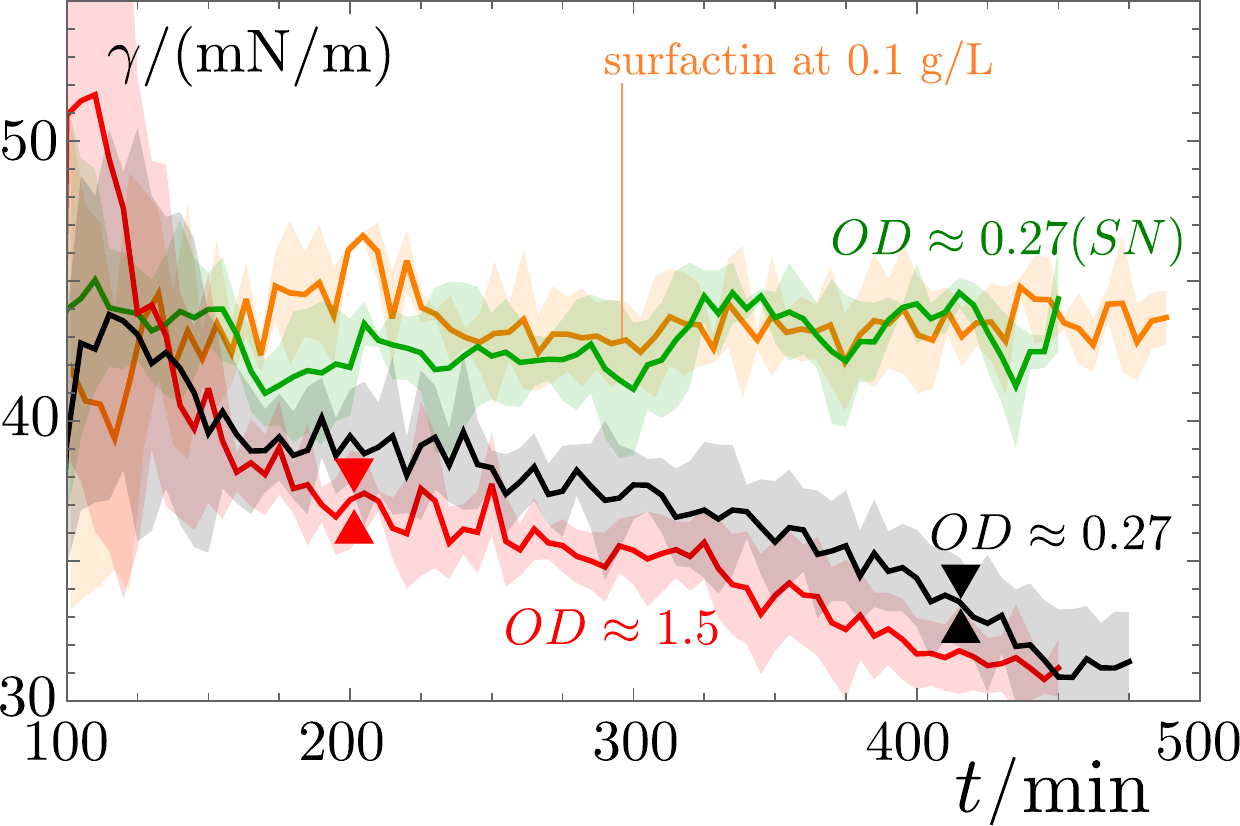}
\caption{The surface tension decreases monotonically over the whole
  duration of the experiment for drops containing \surfactin{}
  producing bacteria. This decrease is not affected by the onset of
  colony surfing at $t=\SI{205}{\minute}$ for the drop of initial
  $\OD=1.5$, $t=\SI{415}{\minute}$ for $\OD=0.27$ (triangles). By
  contrast if the bacterial suspension is centrifuged and only the
  supernatant is deposited on the gel, the surface tension quickly
  becomes constant after the initial decrease due to the \surfactin{}
  produced prior to the removal of the bacteria. A similar trend is
  observed for drops containing only commercial surfactin.}
\label{fig:surfacetension}
\end{figure}

\begin{figure}
\centering
\includegraphics[width=.7\columnwidth]{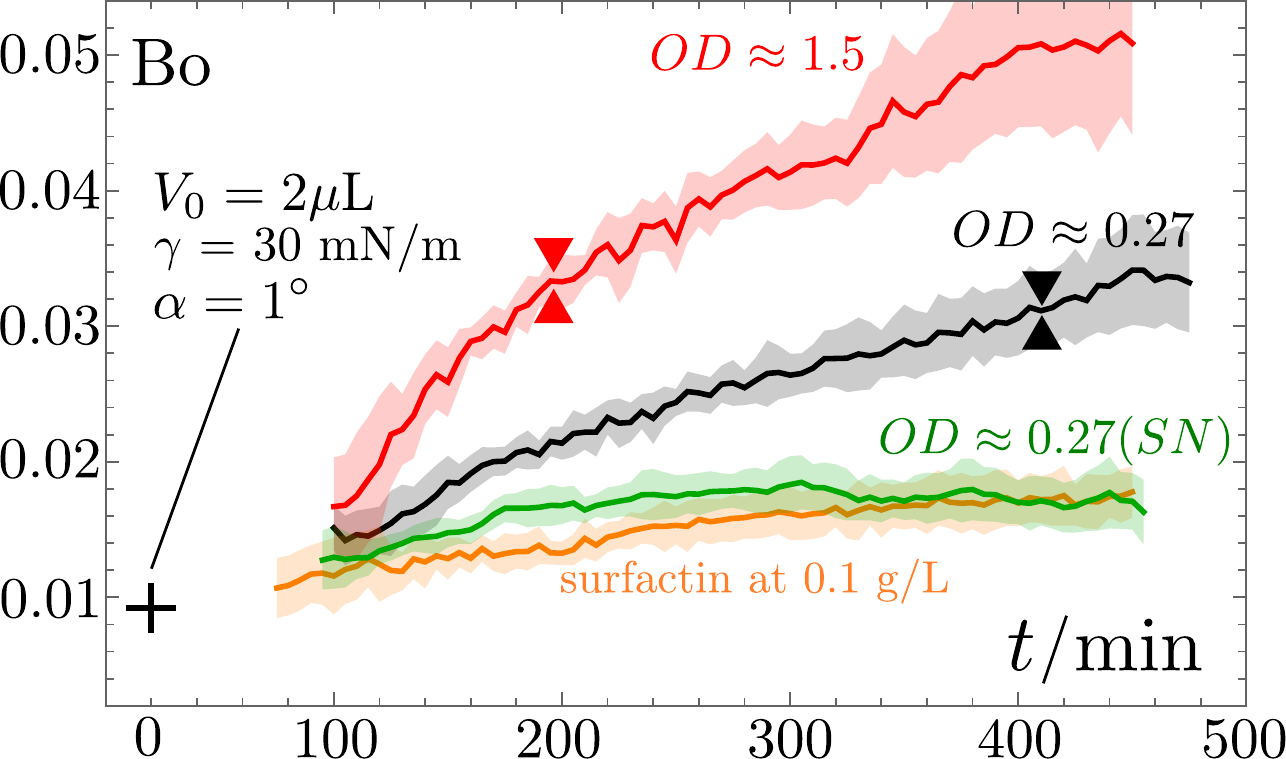}
\caption{The Bond number calculated using the measured volume and
  surface tension (Figs.~\ref{fig:volume} and
  \ref{fig:surfacetension}) exhibits a monotonically increase for
  \surfactin{} producing strains, with a critical value for the onset
  of surfing (triangles) at $\Bond_c^{\bact} \approx 0.03-0.05$. The
  supernatant solution (SN) from an $\OD \approx 0.27$ suspension, and
  the commercial surfactin droplets (\SI{0.1}{\gram\per\litre}) do not
  reach this value and no surfing is observed.}%
\label{fig:bond}
\end{figure}

\subsection*{Lowering of the surface tension}
The second mechanism that \Bsubtilis{} is able to harness directly
impacts the capillary forces through a surfactant-induced reduction of
surface tension. Thanks to the precision of our profilometric data, we
can extract the curvature of the drops at different heights and infer
the drop's surface tension \textit{in situ} during the experiment.
This shows that the constantly produced \surfactin{} molecules lower
the drop's surface tension by almost a factor 2
(Fig.~\ref{fig:surfacetension}). This decrease of surface tension
generates a further increase of the Bond number
(Eq.~\ref{eq:mynameisJames}) by a similar factor.

We show in Fig.~\ref{fig:bond} the overall variation of the Bond
number of \SI{2}{\micro\litre} drops with different initial
concentrations of bacteria. The Bond number increases significantly
over time for high initial optical densities,
$\OD \approx 0.2$ or higher, reaching about ten times its initial
value. Colony surfing (onset marked by triangle symbols) is observed
for a threshold value of
$\Bond_c^{\bact} \approx \mbox{\numrange{0.03}{0.05}}$, a value that
is never reached by the supernatant solution or drops showing even
less volume increase. This establishes a relation between the
\surfactin{} production, volume increase and the possibility of
gravitationally driven onset of motion. Note, however, that the Bond
number at which colony surfing is seen remains one order of magnitude
below the critical Bond number of water which shows that the volume
increase and the reduction of surface tension are only part of the
toolkit available to \Bsubtilis{} for depinning its containing drop.

\subsection*{Enhancing the substrate wettability}
Indeed, the production of surfactin also strongly enhances the
wettability of the agar gel. The surfactant ring (SR) extending around
drops containing surfactants marks the outer limit of the region of
enhanced wettability (Figs.~\ref{fig:kinetics+Bond}
and~\ref{fig:inducedsurf}). Within the SR, contact angles of deposited
water drops are dramatically lowered~\cite{leclere2006}. Indeed when
we place a large drop of water (\SI{30}{\micro\litre}), whose Bond
number is below $\Bond_{\rm c}^{\water}$, next to a colony of
non-motile, \surfactin-producing bacteria (strain OMG\,954), the water
drop first remains immobile. When reached by the SR, it however starts
sliding in the region of larger wettability (Supplementary
Fig.~\SUPPwatercontaminationSR{} and movie~\SUPPmovieInducedSR).

To quantify the capacity of the agar gel surface to \emph{pin} drops,
we extracted the advancing and receding contact angles of sliding
bacterial drops from our profilometric data. For a
$V = \SI{15}{\micro\litre}$ drop, of width
$w = \SI{12.5}{\milli\metre}$, we obtain $\theta_A \approx \ang{6.5}$
and $\theta_R \approx \ang{1.9}$, corresponding to a slope of
\num{0.114} and \num{0.033}, respectively (Supplementary
Fig.~\SUPPcontactangles). The capillary pinning force which opposes
the gravitational pull is directly proportional to
$\Delta\cos\theta = \cos \theta_R - \cos \theta_A$. Drop sliding is
thus only possible when the gravitational pull overcomes a critical
value~\cite{dussan1983}:
\begin{equation}
\rho V g \sin \alpha = w \gamma (\cos\theta_R - \cos\theta_A).
\label{eq:dussan}
\end{equation}
This is equivalent to saying that the critical Bond number is
$\Bond_c = \frac{w}{V^{1/3}} \Delta\cos\theta\simeq 0.03$ which is
very close to the measured Bond number at which colony surfing is
observed in our experiments (Fig.~\ref{fig:bond}). This third
mechanism thus concludes our discussion on the depinning of
\Bsubtilis{} droplets seen in our experiments by providing a
quantitative match with the corresponding critical Bond number.

\subsection*{Surfactin drops}
To test more quantitatively the role of \surfactin{}, we quantified
its concentration in the supernatant of bacterial suspensions through
High Performance Liquid Chromatography (HPLC, see \textit{Materials
  and Methods} and suppl. Fig.~\SUPPhplccalibration{} and
Table~\SUPPhplcconcentrations). For a liquid culture of the
\surfactantstrain{} strain at $\OD=0.27$, we measured a surfactin
concentration of \SI{106+-32}{\milli\gram\per\litre}. We then compared
the behaviour of a droplet of the supernatant of this bacterial
suspensions with a drop of commercial \surfactin{} in water at a
concentration of \SI{100}{\milli\gram\per\litre}. We show in
Fig~\ref{fig:volume},~\ref{fig:surfacetension}, and~\ref{fig:bond}
that their evolutions are very similar. Together with the experiments
on the surfactin deficient strain 168, this shows surfactin to be the
main tool used by \Bsubtilis{} to depin the droplets. Note that even
higher concentrations of \surfactin{} can even induce a depinning of
the contact line. This however does not lead to a sustained motion of
the droplet due to the finite extent of the surfactin ring in the
absence of continuous production of surfactin by bacteria inside the
droplet (Suppl. movie~\SUPPmovieSurfactinSlide).

\section*{Discussion} 
In this article, we have shown how \Bsubtilis{} colonies can modify
their environment, creating a mobile drop which propels them across
solid substrates at high velocities (\SI{3.1}{\centi\metre\per\hour}
on a \ang{1} slope in supplementary movie \SUPPmovieInducedWater,
potentially much more on steeper slopes), faster than any other
surface-bound translocation mechanism described so far. Our
measurements reveal that the bacteria have at least three distinct
physico-chemical mechanisms they can use to this end: Liquid is
extracted from the environment, inflating the drop and increasing the
gravitational pull, the surface tension is lowered, and the
wettability of the environment is significantly increased. This leads
to a situation where the bacterial drop starts to run downhill on very
shallow slopes, as low as \ang{0.1}, under a gravitational pull which
is two orders of magnitude too weak to depin a similar drop of water.
While the influence of surfactant production on bacterial
translocation has been observed in the context of flagellated
swarming~\cite{julkowska2005,partridge2013,beer2009,henrichsen1972,
  harshey2003, leclere2006}, or on biofilm spreading in liquid
culture~\cite{angelini2009}, this is to our knowledge the first
observation of a surfactant-induced translocation mode where
individual motility is not required. Its efficiency, making
\emph{colony surfing} the fastest surface translocation mode, is
therefore even more remarkable, and arguably relevant in nature.
Indeed, colony surfing can also occur if only part of the various
mechanisms we uncovered are at play: a small bacteria-laden drop
suspended on the upper plate of a sealed plastic box containing a
water reservoir can undergo sliding while water droplets of similar
sizes would remain stuck by capillary forces (\SUPPpetridishsliding).
Osmotic-induced surfing should thus be a rather generic phenomenon in
humid environment. Furthermore, the exploitation of the
physico-chemistry of surfactants to facilitate wetting and
translocation will also be relevant in habitats with strongly varying
humidity, such as soils~\cite{or2007} where surfactant-producing
bacteria are rather common~\cite{bodour2003distribution}. Capillary
forces are indeed known to control the water flow in porous
media~\cite{sadjadi2013} so that collective surfing could strongly
enhance migration speeds and distances of bacterial colonies through
the soil pore network.

\matmethods{%
  \subsection*{Volume and surface tension measurement}
  \setlength{\intextsep}{0pt}
  \setlength{\columnsep}{1em}
  \begin{wrapfigure}[8]{R}{0pt}%
    \includegraphics[width=0.29\columnwidth]{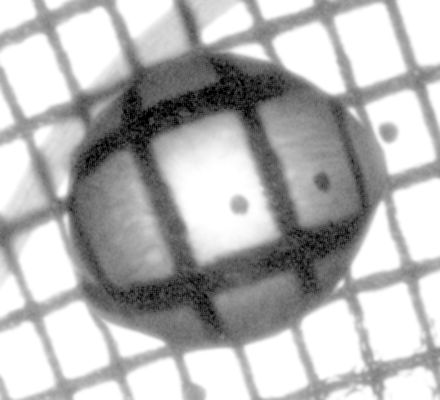}%
  \end{wrapfigure}
  
  \textit{In situ} volume and surface tension measurements are
  performed via a Moir\'e grid method where we calculate the free
  surface deformation from the distortion of the image of a grid
  modulating light intensity, as seen through the
  drop~\cite{banaha2009}. In our set-up the grid translates uniformly,
  so that the intensity at every pixel varies periodically in time. A
  tilt in the air-drop or air-gel interface causes an apparent shift
  of the grid because of light refraction, producing a phase shift in
  the intensity signal. This shift can be precisely measured and
  related to the local interface slope through Snell's law. We use two
  perpendicular grids to measure two orthogonal slopes $h_x(x,y)$ and
  $h_y(x,y)$ simultaneously. The full surface profile $h(x,y)$ is
  obtained by a Fourier-transform integration (Frankot-Chellappa), and
  further integration
  $V = \int_{S} h(x,y) \,{\mathrm d}x \,{\mathrm d}y$ gives the drop
  volume. Conversely the derivatives $h_{xx}$, $h_{xy} = h_{yx}$ and
  $h_{yy}$ yield the mean interface curvature $\kappa$:
  \begin{equation*}
    \kappa = \frac{h_{xx}(1+h_y^2) + h_{yy}(1+h_x^2)
      - 2h_{xy} h_x h_y}{(1+h_x^2 + h_y^2)^{3/2}}.
  \end{equation*}
  This curvature gives rise to a pressure jump across the air-drop
  interface, which in our very slow drops ($\Capillary < 10^{-6}$)
  depends only on height because hydrostatic pressure gradients in-
  and outside the drop are different~\cite{younglaplace}:
  \begin{equation*}
    \gamma \kappa(x,y) = p_0 + \rho g h(x,y).
  \end{equation*}
  With $p_0$ a constant, $\rho$ essentially the density of water and
  the gravitational acceleration $g$ known, the surface tension
  $\gamma$ is immediately obtained from the slope of a linear fit of
  our measured interface height vs. curvature data. Calibration of the
  setup and subsequent algorithms is achieved by comparing the
  determined values for volume $V$, apex height $h^{\text{max}}$ and
  curvature $\kappa$ of various spherical-cap lenses with the
  manufacturer's specifications. Deviations are found to be within a
  \SI{1}{\percent} margin. Furthermore, surface tension measurements
  on water, glycerol and surfactin solution were found to agree with
  measurements through the pendent drop method~\cite{daerr2016jors} to
  within \SI{2}{\percent}.

  \subsection*{Contact-angle determination}
  The contact line position is determined by noting that the second
  derivatives $h_{xx}$ and $h_{yy}$ of the height profile $h(x,y)$
  must have a local maximum there, corresponding to the discontinuity
  at the triple liquid-solid-air interface (Fig.~\SUPPcontactangles).
  This yields the closed contact line around the drop with a precision
  of a few pixels, or about \SI{1}{\percent} of a drop diameter of
  \SIrange{5}{15}{\milli\metre} at an image scale of
  \SI{41.1}{px/\milli\metre}. The determined contact-angle
  measurements are again calibrated by comparison with the known
  border-gradients $h_{x,y}^{\text{max}}$ of spherical cap lenses and
  show good agreement within a \SI{5}{\percent} range.

  \subsection*{Agar gel preparation}
  A \SI{1.4}{\percent} (mass) stock solution is prepared from granular
  Agar (Difco). Just before an experiment, this solution is heated by
  microwave radiation until reaching the boiling point of water. The
  molten gel is weighted and solvent-loss due to evaporation is
  compensated for. We mix it with a liquid synthetic nutrient
  B-medium~\cite{antelmann1997} (prepared at twice the final
  concentration) at a 1:1 ratio to obtain a \SI{0.7}{\percent} gel
  solution with nutrients, from which \SI{25}{\milli\litre} are poured
  in a Petri dish whose lid is closed immediately afterwards.
  Subsequent gelling takes place at constant temperature
  (\SI{30}{\celsius}) and on a flat surface for \SI{20}{\minute}. The
  lid is then removed for a duration $t_d$ (\SI{1}{\minute} unless
  specified otherwise) to dry the solidified gel surface in ambient
  air (RH=\SI{35}{\percent}). The process of drop deposition and
  transfer may add up to one minute to this open-lid drying time.

  \subsection*{Bacterial strains and preparation}
  The \Bsubtilis{} strains (Laboratory strain 168, genotype
  \textit{trpC2 swrA sfp\textsuperscript{-}}; \surfactantstrain, which
  is strain 168 restored to \textit{sfp\textsuperscript{+}} on the
  \textit{thrC} locus, also known as OMG\,930~\cite{hamze2009};
  OMG\,954, non-flagellated by deletion of the \textit{hag} gene from
  \surfactantstrain~\cite{hamze2009}) are taken from a
  \SI{-80}{\celsius} storage, spread out on a solid \SI{0.7}{\percent}
  Agar medium with LB broth and antibiotics (\surfactantstrain{}:
  erythromycine \SI{0.5}{\micro\gram \per \milli\litre}, lincomycin
  \SI{12.5}{\micro\gram \per \milli\litre}; in the case of OMG\,954
  additionally chloramphenicol \SI{5}{\micro\gram \per \milli\litre}),
  and incubated at \SI{30}{\celsius} and \SI{70}{\percent RH} for
  \SI{24}{\hour}. Microcolonies are taken from this gel and grown
  overnight in liquid B-antibiotics culture at \SI{37}{\celsius}. The
  suspension is then diluted to an optical density ($\OD$ at
  \SI{600}{\nano\metre}) of \num{0.1}, and after \SI{1}{\hour} a
  second time to $\OD = \num{0.1}$. It is left in the shaker until
  bacterial growth saturates. The doubling time in B medium at
  \SI{30}{\celsius} resp. \SI{37}{\celsius} is around
  \SI{100}{\minute} resp. \SI{85}{\minute} for strain 168, and
  \SI{110}{\minute} resp. \SI{90}{\minute} for \surfactantstrain{}
  (Fig.~{\SUPPopticaldensity}). \SIrange{2}{4}{\hour} after departing
  from exponential growth, \SI{2}{\micro\litre} drops of suspension
  (diluted to the desired \OD) are deposited on the previously
  prepared agar gel and transfered into a constant climate chamber
  (Memmert HPP 110) at controlled humidity and at a temperature of
  \SI{30}{\celsius}. An optical density of $\OD=0.27$ corresponds to
  \SI{1e8}{cell\per\milli\litre}, i.e. a volume fraction
  $\Phi\simeq \SI{0.05}{\percent}$.

  \subsection*{Determination of surfactin concentration by HPLC}

  Concentrations of surfactin in surfactin-producing bacteria
  suspensions sampled from bulk and droplets deposited on gel were
  determined by reverse-phase $\text{C}_{18}$ HPLC on a Varian Prostar
  equiped with a UV 320 detector by reference to a commercial
  surfactin standard (Surfactin from \textit{B. subtilis},
  \SI{98}{\percent} pure, Sigma-Aldrich, batch n° 086K4109). The
  culture medium, as well as supernatant from a liquid culture of
  strain 168 which does not produce surfactin were also analysed as
  controls.

  {\it Acquisition of calibration data.} A \SI{1.3}{\gram\per\litre}
  mother solution of commercial surfactin was prepared by dissolving
  \SI{1.3}{\milli\gram} surfactin in \SI{1}{\milli\litre} Ethanol
  (\SI{95}{\percent}). The mother solution was then diluted in MilliQ
  water to prepare calibration standards with \num{10}, \num{100},
  \num{200}, \SI{300}{\milli\gram\per\litre} concentrations.
  \SI{100}{\micro\litre} samples were injected and eluted during
  \SI{30}{\minute} using a mobile phase made of \ce{ACN}/\ce{H2O}
  80:20 (v/v) supplemented with \SI{0.1}{\percent (v)} of
  trifluoroacetic acid, at a flow rate of
  \SI{1}{\milli\litre\per\minute}. Detection was performed by
  measuring the absorbance at \SI{205}{\nano\metre}.

  {\it Acquisition of sample data.} Samples were filtrated on
  \SI{0.2}{\micro\metre} cellulose acetate syringe filters, injected
  as \SI{20}{\micro\litre} aliquots and eluted using the settings
  described for calibration standards.

  {\it Building of the calibration model and estimation of
    concentration in samples.} Surfactin exists in several isomeric
  forms that usually coexist in the same extract. The proportions of
  the various isomers vary from one bacterial species to the other
  (Fig.~\SUPPhplccomparison). Hence, for calibration, a global
  surfactin-correlated signal was built by summing the contributions
  of several isomers to the chromatogram. The peaks at \num{14},
  \num{18}, \num{19} and \SI{23.5}{\minute} were fitted to Gaussian
  curves, integrated and summed to give rise to a total surfactin
  signal that was plotted against the concentrations of the surfactin
  standards (Fig.\,\SUPPhplccalibration). A linear calibration model
  was built using the standards up to \SI{300}{\milli\gram\per\litre}.
  It had a multiple R-squared correlation coefficient of \num{0.985}
  with a residual standard error of \num{0.0026} on \num{5} degrees of
  freedom. The limit of detection (LOD) was calculated with the usual
  definition of the concentration corresponding to the signal of the
  blank sample plus three standard deviation of the blank. The LOD was
  \SI{44}{\milli\gram\per\litre}. Next, the equivalent concentrations
  of surfactin in the samples and controls were determined from the
  total surfactin-correlated peak areas of the corresponding
  chromatograms. }

\showmatmethods 

\acknow{We thank J.-F. Berret, M. Costalonga, C. Cottin-Bizonne, B.
  Desmazi{\`e}res, L. Hamouche, K. Hamze, I.~B. Holland, S. S{\'e}ror
  and M. Zhao for fruitful discussions. We thank David Clainquart and
  Marie-Evelyne Pinart from the Chemistry department of the University
  Paris Diderot for their assistance on the HPLC platform. This
  research was funded through the ANR grant \emph{Bactterns}, CNRS
  grant PEPS ``Micromanipulation de particules actives'' and CNRS
  grant PIR ``Mécanismes physiques et biologiques de la migration en
  masse de B. subtilis''.}

\showacknow 

\pnasbreak

\bibliography{biblio}


\pnasbreak

\setlength\parindent{0pt}

\newcommand{\movieonline}[1]{#1}
\newcommand{\movielink}[1]{%
    \begingroup\edef\x{
      \lowercase{\endgroup
        \def\noexpand\foo{#1}}}%
    \x
    \href{http://www.msc.univ-paris-diderot.fr/~daerr/research/colonysurfingmovies.html\#\foo}{\includegraphics[width=0.5\linewidth]{movies/#1_still}}%
}

\newcommand{\SImovie}[2]{{\centering \movielink{#1}\smallskip\\}%
  \textbf{Movie #1}: #2\bigskip\\\strut}
\newcommand{\SIfigure}[4][]{{\centering \includegraphics[#1]{#2}\smallskip\\}%
  \textbf{Figure #3}: #4\bigskip}
\newcommand{\SItable}[3]{{\centering #2\smallskip\\}%
  \textbf{Table #1}: #3\bigskip}

\section*{Supplementary information}


\movieonline{Movies can be viewed online at
  \url{http://www.msc.univ-paris-diderot.fr/~daerr/research/colonysurfingmovies.html}
  or at
  \url{http://www.pnas.org/content/suppl/2017/05/23/1703997114.DCSupplemental};
  click on the still pictures to open movie. \bigskip}

\SImovie{\SUPPmovieSurf}{Movie of a colony surfing event of a
  $\SI{2}{\micro\litre}$ drop ($\OD=0.27$) on a \ang{1} slope
  (corresponding to main article figure~1). Time is indicated in
  hours:minutes, the time interval between two frames is $5\,$min.}

\SImovie{\SUPPmovieInducedSN}{Movie showing induced drop sliding of a
  $\SI{10}{\micro\litre}$ \emph{supernatant} drop (right) contaminated
  by a dendrite of swarming bacteria from a $\SI{1}{\micro\litre}$
  drop of bacteria ($\OD = 1.5$, left) at $t\simeq 10\,$h. The
  inclination angle is \ang{1}. See main article figure~3 for details.
  Time is indicated in hours:minutes, the time interval between two
  frames is $5\,$min.}

\SImovie{\SUPPmovieInducedWater}{Movie showing induced drop sliding of
  a $\SI{50}{\micro\litre}$ \emph{water} drop (left) contaminated by a
  dendrite of swarming bacteria from a $\SI{30}{\micro\litre}$ drop of
  bacteria (right) at $t\simeq 110$\,min. Here, to reduce initial drop
  spreading, the initial amount of \surfactin{} present in the drop
  was minimised through the following procedure: the bacterial
  suspension was grown to $\OD = 1.5$ and centrifuged; the supernatant
  was removed and the bacteria resuspended in fresh nutrients (final
  $\OD = 1.5$) before deposition. The inclination angle is \ang{1}.
  See supplementary figure~\SUPPwatercontamination{} for details. Time
  is indicated in hours:minutes, the time interval between two frames
  is $2.5\,$min.}

\SImovie{\SUPPmovieInducedSR}{Movie showing how a
  $\SI{30}{\micro\litre}$ water drop (left) starts to slide after
  being reached by the \surfactin{} ring emitted by a
  $\SI{30}{\micro\litre}$ drop of non-motile bacteria (right). As for
  movie~\SUPPmovieInducedWater, to reduce initial drop spreading, the
  initial amount of \surfactin{} present in the drop was minimised
  through the following procedure: the bacterial suspension was grown
  to $\OD = 1.5$ and centrifuged; the supernatant was removed and the
  bacteria resuspended in fresh nutrients (final $\OD = 1.5$) before
  deposition. See supplementary figure~\SUPPwatercontaminationSR{} for
  details. Time is indicated in hours:minutes, the time interval
  between two frames is $2.5\,$min. }

\SImovie{\SUPPmovieSurfactinSlide}{Movie showing the evolution of a
  $\SI{2}{\micro\litre}$ drop of commercial \surfactin{} solution
  (\SI{0.5}{\gram\per\litre}) on the gel surface. As for bacterial
  suspensions, we observe a volume increase, and the outward
  propagation of a \surfactin{} ring. The wetting hysteresis is
  reduced to
  $\Delta\cos\theta = \cos \theta_R - \cos \theta_A = 0.003$ at the
  onset of motion. As opposed to the drops of bacterial suspension the
  drop moves only through a finite distance, probably because
  sustained \surfactin{} production is required to continuously modify
  the wetting properties of the gel ahead of the sliding drop. Time is
  indicated in hours:minutes, the time interval between two frames is
  \SI{5}{\minute}.}


\SIfigure[width=\columnwidth,height=0.77\textheight,keepaspectratio]{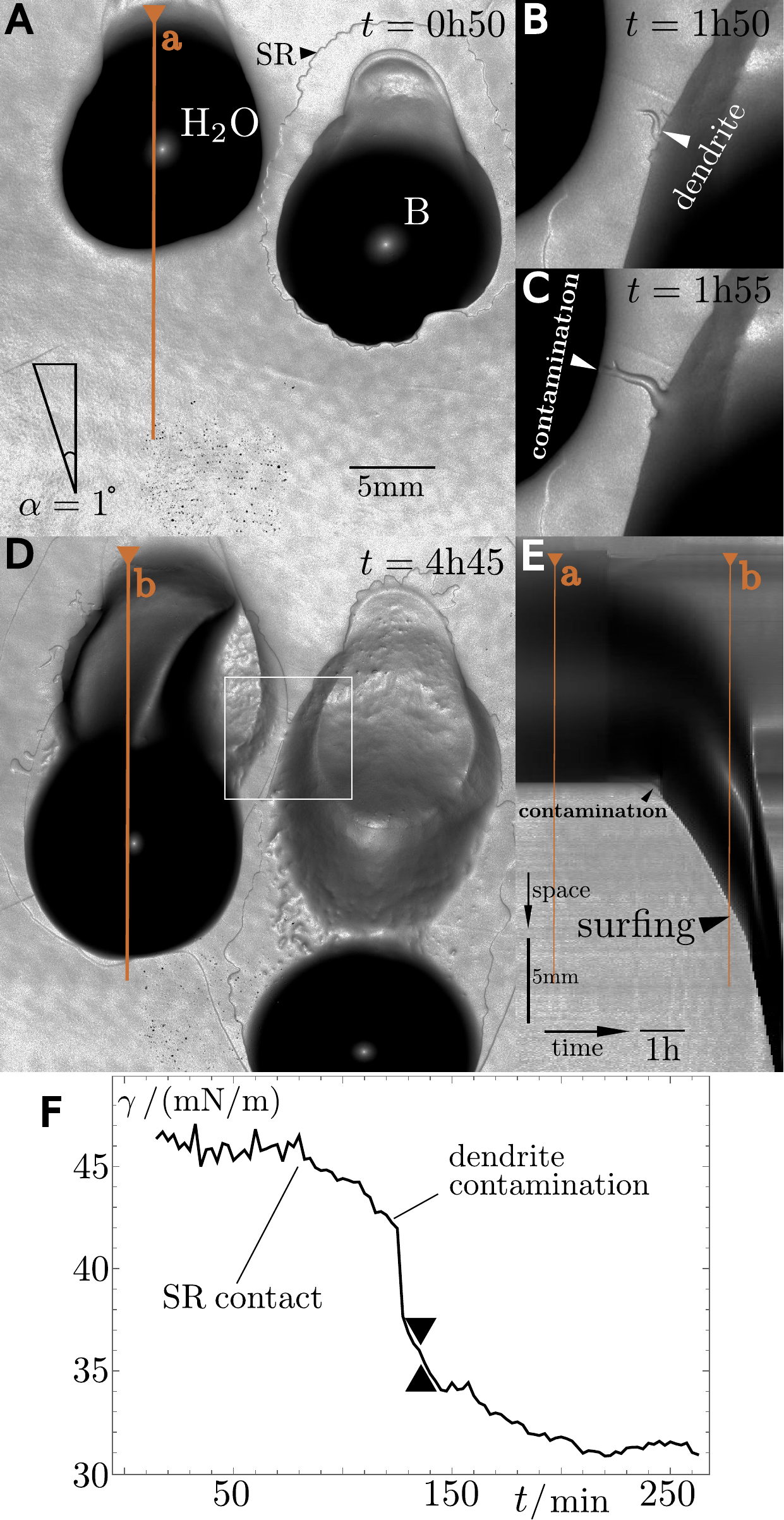}%
{\SUPPwatercontamination}{%
  \textbf{(A-D)} Snapshots of induced colony surfing, where the
  grey-level indicates the local slope. The right drop is a
  \SI{30}{\micro\litre} drop of bacterial suspension ($\OD=1.5$),
  while the left drop is \SI{50}{\micro\litre} of distilled water. The
  water drop starts to slide upon being contaminated by a dendrite of
  swarming bacteria (detail at $\SI{1}{\hour}\,50$, white rectangle
  region). \textbf{(E)} Kymograph of a cut through the water drop
  along the direction of motion showing the onset of sliding
  immediately after the contamination (black triangle). \textbf{(F)}
  Surface tension of the water drop. The surface tension starts
  decreasing from the moment the drop is reached by the
  \surfactin{}-ring, indicative of surfactant molecules populating the
  interface. A much sharper decrease is associated with the
  contamination of the drop by a dendrite of migrating bacteria.}

\SIfigure[width=0.85\columnwidth]{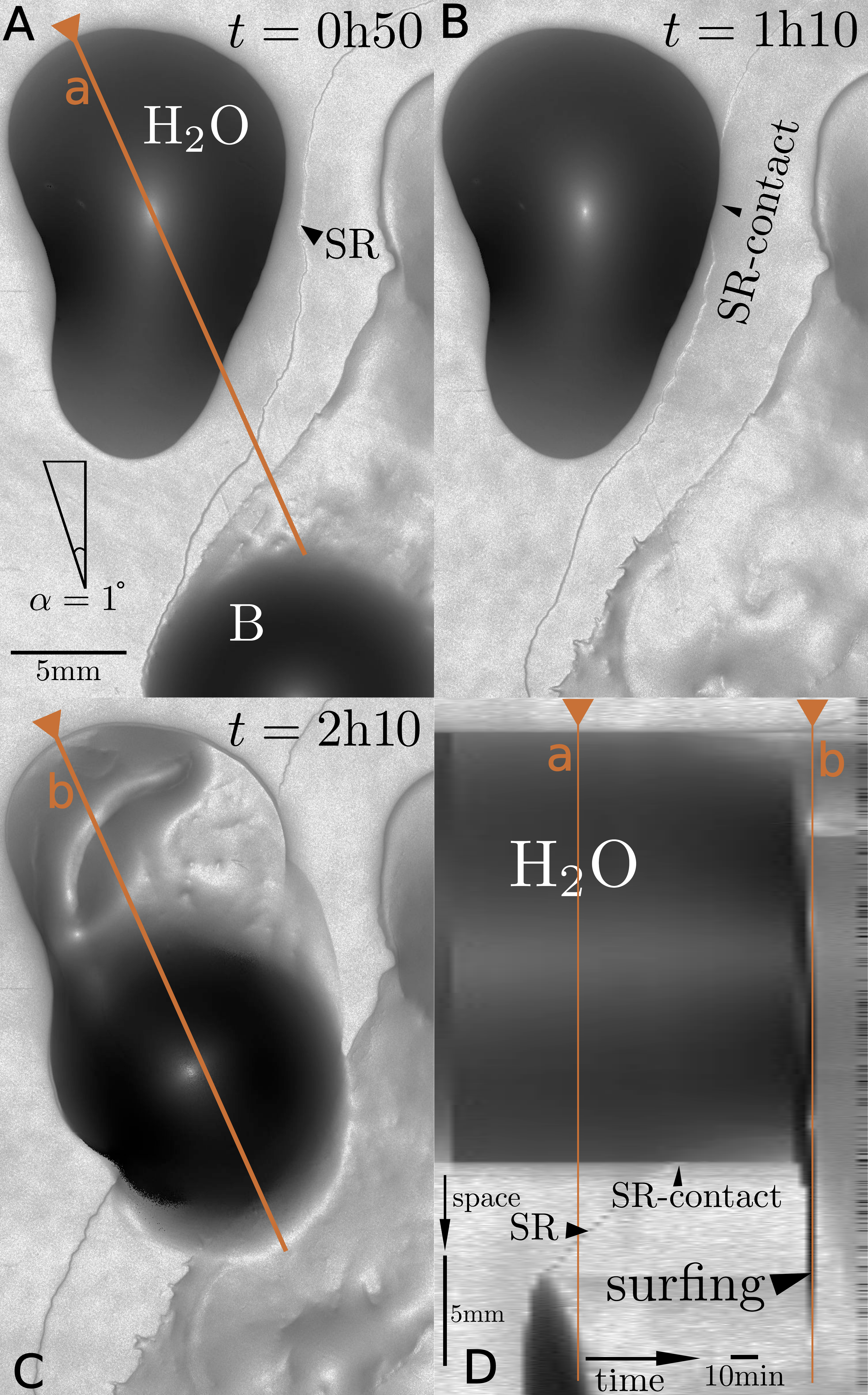}{\SUPPwatercontaminationSR}{%
  \textbf{(A-C)} Snapshots of induced colony surfing, where the
  grey-level indicates the local slope. The right drop is a
  \SI{30}{\micro\litre} drop of bacterial suspension ($\OD=1.5$),
  while the left drop is \SI{30}{\micro\litre} of distilled water. The
  water drops starts to slide some time after the surfactant ring SR
  issued by the bacteria-laden drop reaches it, $t=1 \text{h}50$.
  \textbf{(D)} Kymograph of a cut through the water drop along the
  direction of motion showing the sudden onset of sliding around
  \SI{40}{\minute} after contact with the SR.}

\SIfigure[width=0.8\columnwidth]{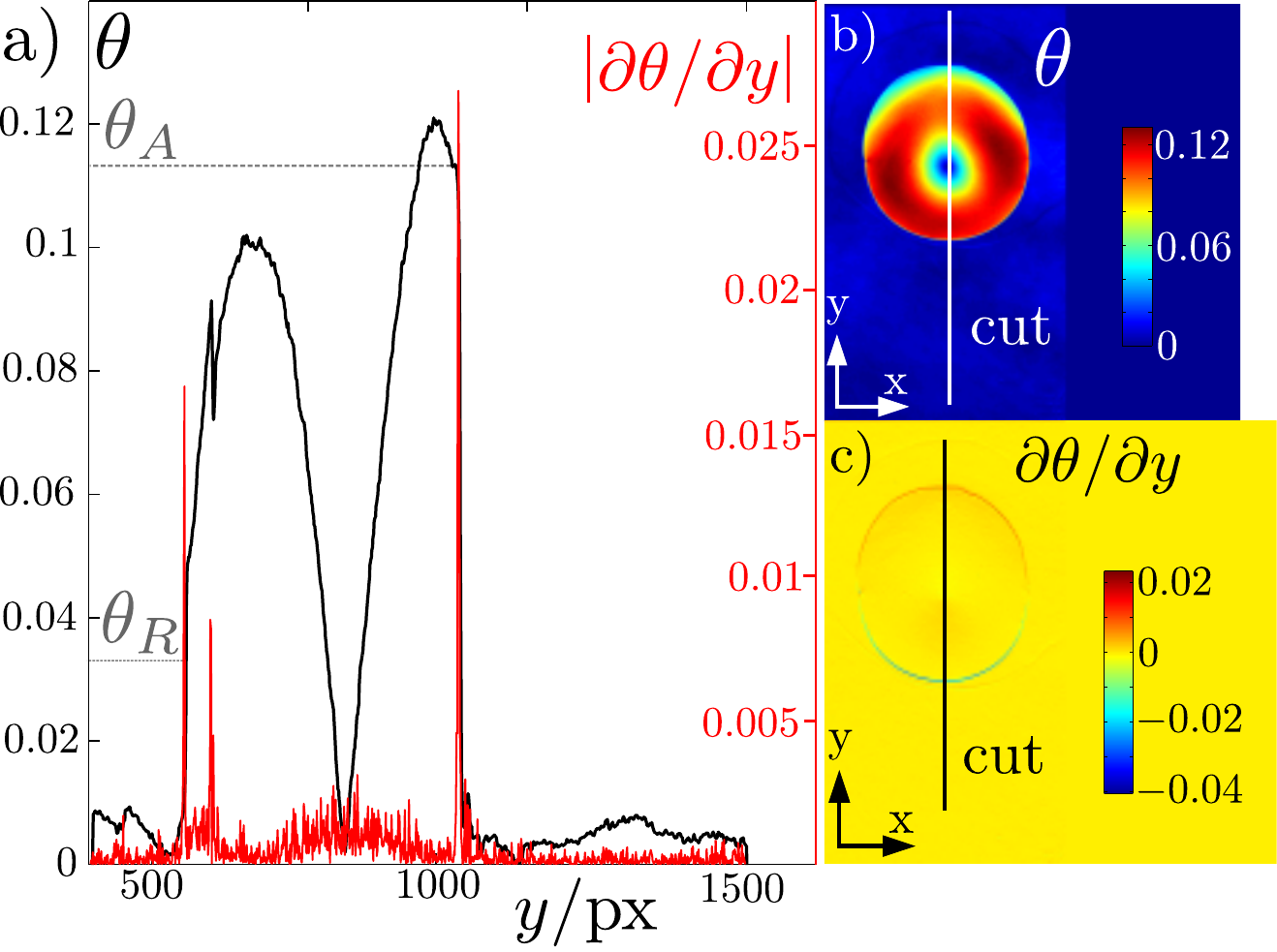}{\SUPPcontactangles}{Local
  contact line position and angles for a drop at the onset of colony
  surfing (corresponding to main article fig.~1). \textbf{(a)}: The
  advancing and receeding contact angles are determined as the
  intersection of the linear extrapolation of the free surface slope
  $\theta(y)$ to the position of the contact line given by the local
  maxima of the red curve. Here
  $\Delta\cos\theta = \cos \theta_R - \cos \theta_A \simeq 0.006$.
  \textbf{(b)} and \textbf{(c)}: Snapshot of the local angle of the
  free drop surface and of its derivative
  $\partial \theta / \partial y$ (top view).}

\SIfigure[width=0.7\columnwidth]{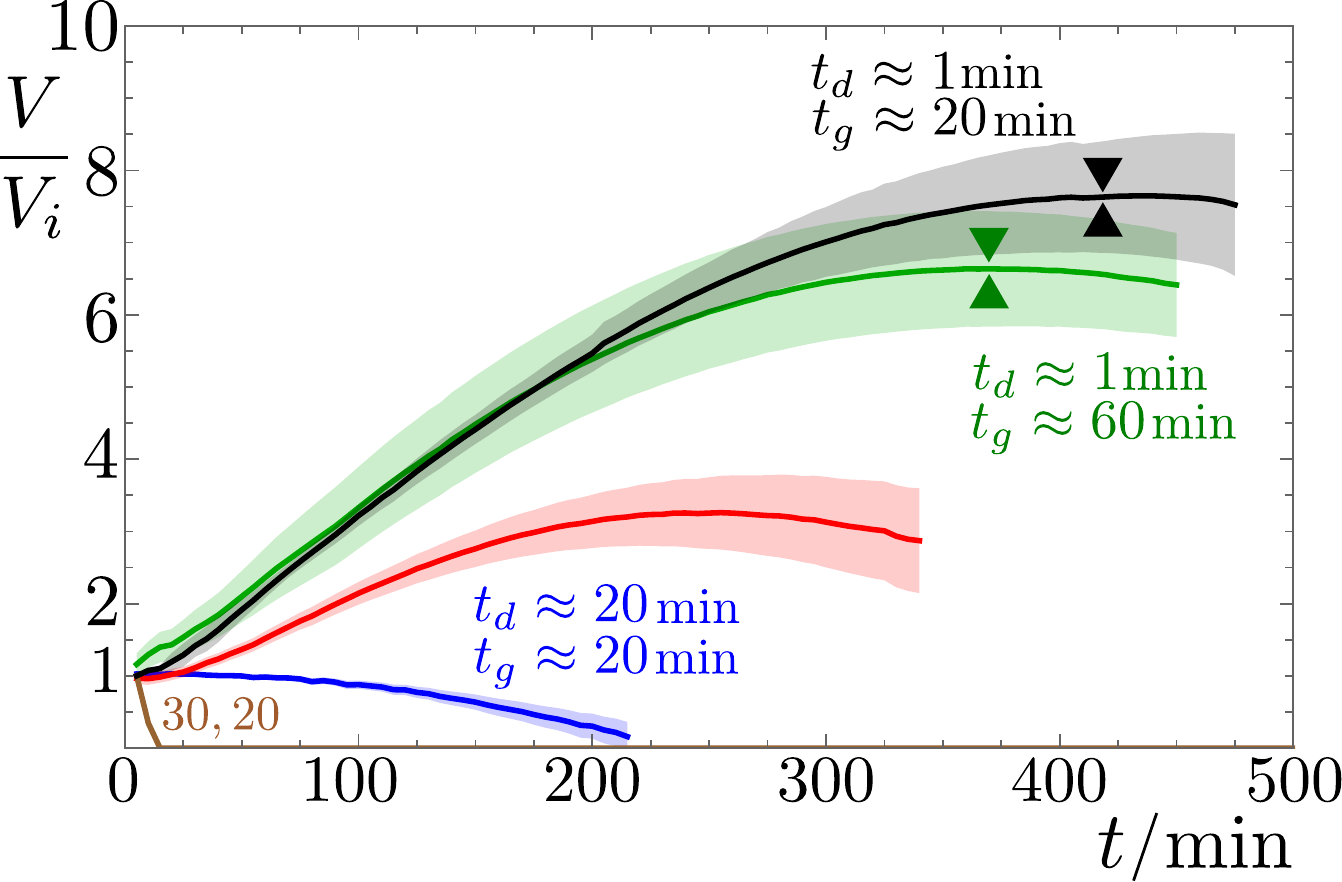}{\SUPPdryingtime}{Impact
  of gelling and drying time $t_g, t_d$ on the inflation of deposited
  drops ($\OD = 0.27$, $V_i=\SI{2}{\micro\litre}$) measured by the
  relative volume increase $V/V_i$. On gels dried for
  \SI{10}{\minute}, the volume gain $V/V_i$ remains well below
  \num{4}, half of the maximum volume gain for the shortest drying
  times. For drying times larger than about \SI{20}{\minute}, the
  volume of deposited drops monotonically decreases. A longer
  \emph{gelling} time (tripled from \SI{20}{\minute} to
  \SI{60}{\minute}) in turn has no significant effect on the evolution
  of deposited drops. Centerlines in saturated colors represent the
  mean value of a series of measurements, with the shaded regions
  around the curves indicating the standard deviation. Triangles mark
  the onset of colony surfing.}

\SIfigure[width=0.7\columnwidth]{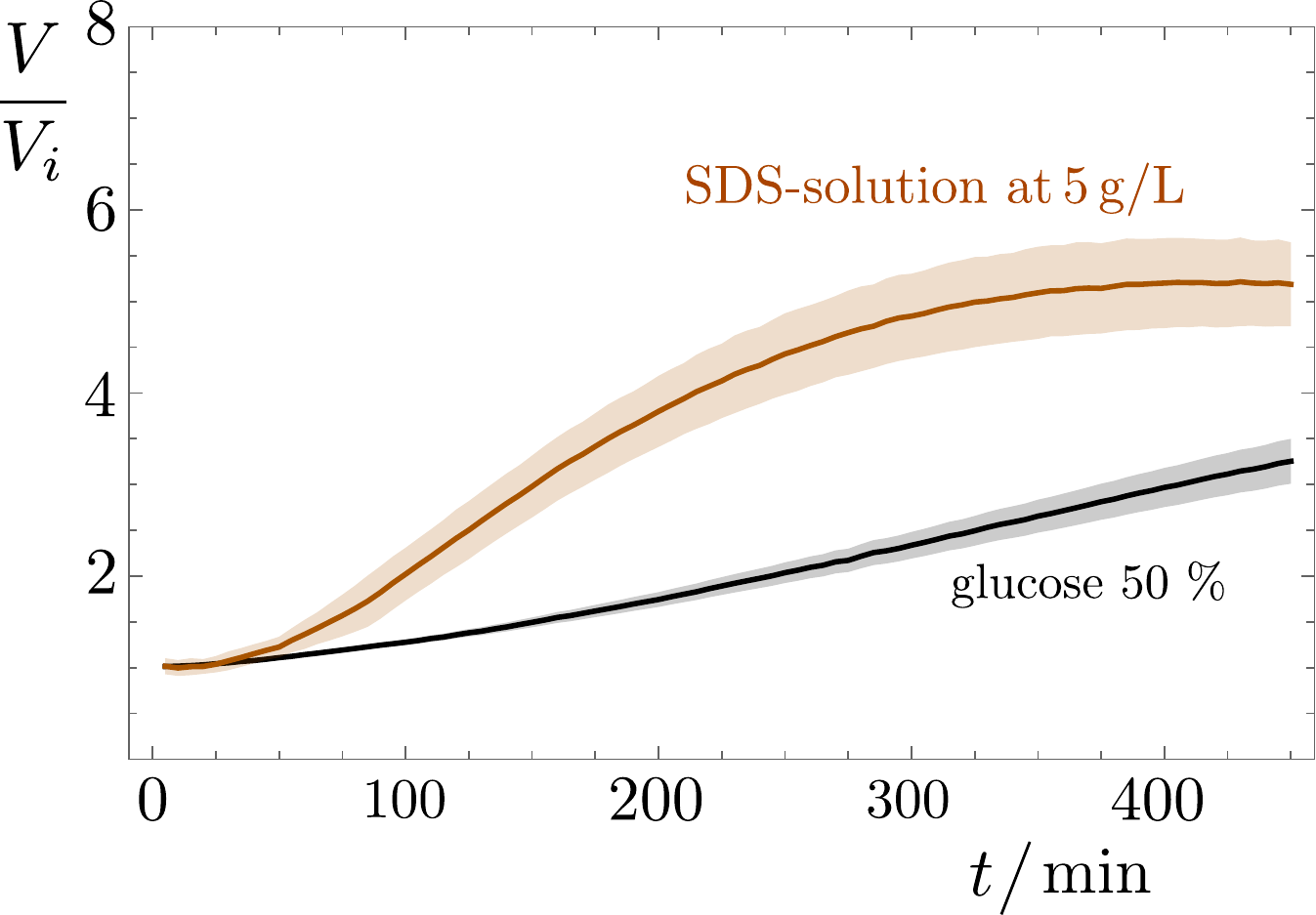}{\SUPPglucSDS}{Evolution
  of the volume of a glucose (50\,\% v/v) resp. a sodium dodecyl
  sulphate (SDS 5\,g/l) drop at the gel surface after deposition at
  $t=0$. The middle line in saturated color is the mean over several
  runs, the standard deviation being shown as shaded bands around the
  mean.}

\SIfigure[width=0.7\columnwidth]{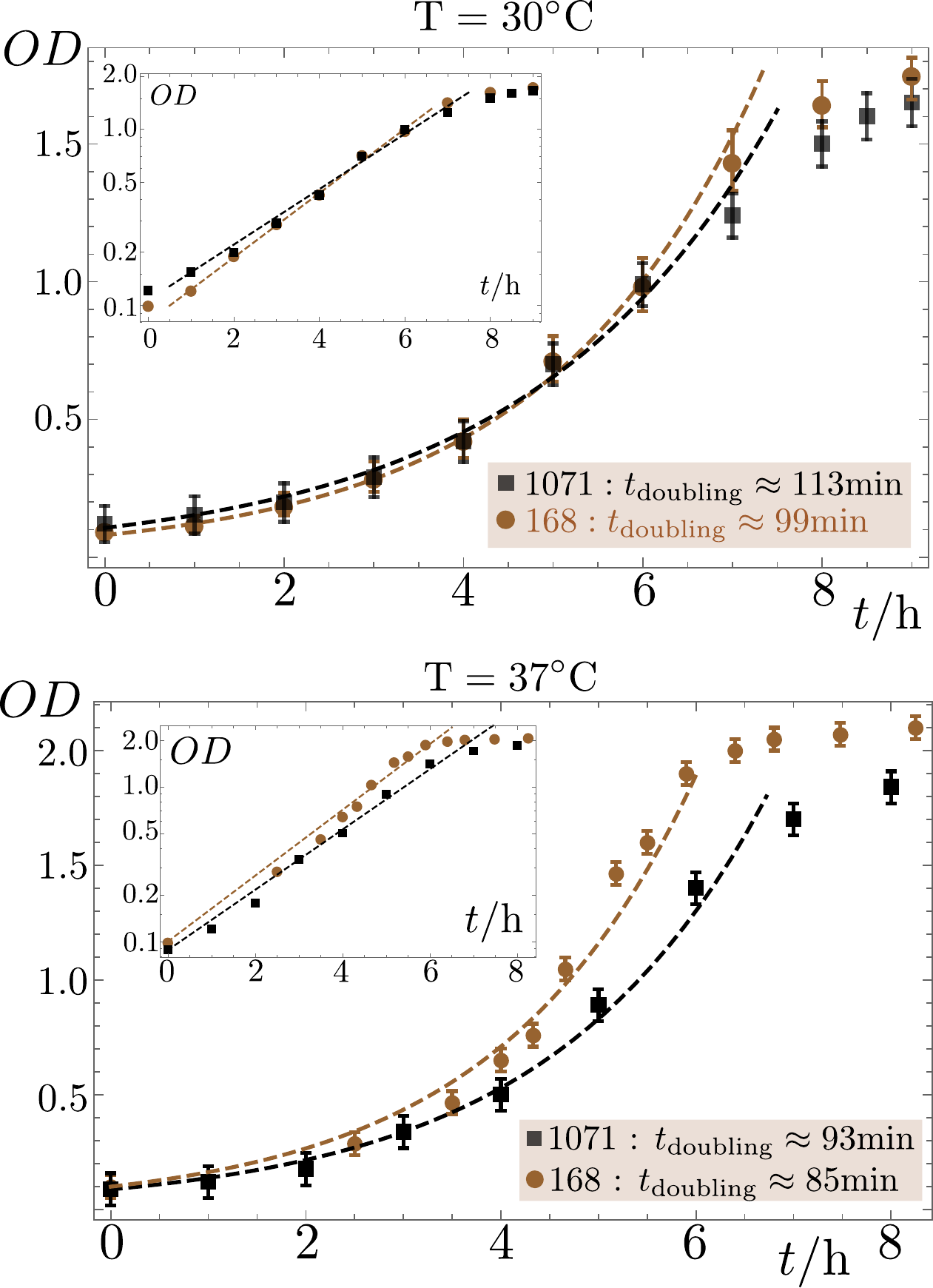}{\SUPPopticaldensity}{The
  doubling time for bacteria grown with aeration in an agitated liquid
  B medium is estimated from optical density (\OD) measurements at a
  wavelength of \SI{600}{\nano\metre}. At a temperature of
  \SI{30}{\celsius} resp. \SI{37}{\celsius}, we find around
  \SI{100}{\minute} resp. \SI{85}{\minute} for strain 168, and
  \SI{110}{\minute} resp. \SI{90}{\minute} for \surfactantstrain.}

\SIfigure[width=\columnwidth]{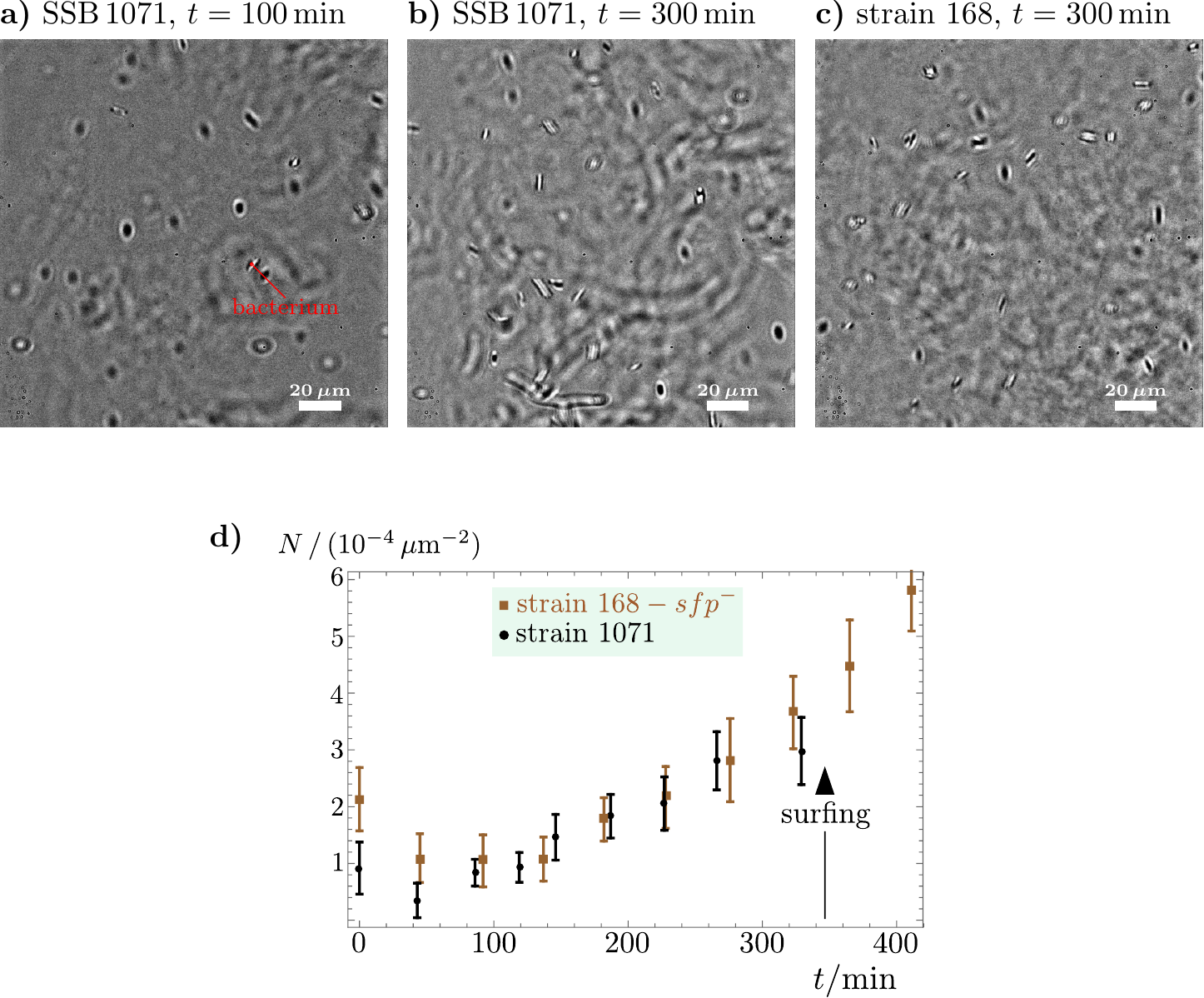}{\SUPPbactdrop}{%
  Monitoring the bacterial concentration in sessile droplets
  (initially: $V = \SI{5}{\micro\litre}$, $\OD=0.27$) on agar
  (\SI{0.7}{\percent}, B medium, $T=\SI{303}{\kelvin}$, RH =
  \SI{90}{\percent}) prior to colony surfing shows that the suspension
  remains dilute over the course of the experiment. \textbf{(a-c)}
  Microscope images at 40x magnification, focussed at mid-height in
  the drop
  ($z=0.5h_{\text{max}}, h_{\text{max}} \simeq
  \SI{0.3}{\milli\metre}$). The depth of field is of the order of
  \SI{1}{\micro\metre}. The bacterial concentration within the droplet
  remains very dilute even at $t=\SI{300}{\minute}$ for both strains
  168 and \surfactantstrain. \textbf{(d)} Mean density of bacteria in
  a slab such as shown in the top row of images (a-c), averaged across
  the droplet's height, i.e. over $\simeq 15$ to $\simeq 20$ slabs ---
  depending on the droplet growth --- recorded every
  \SI{14}{\micro\metre} from top to bottom. The thickness of a slab is
  set by the depth of field, $\simeq \SI{1}{\micro\metre}$. No major
  increase in bacterial density is observed. Note that only strain
  {\surfactantstrain} undergoes colony surfing (at
  $t\simeq\SI{350}{\minute}$). }

\SIfigure[width=0.8\columnwidth]{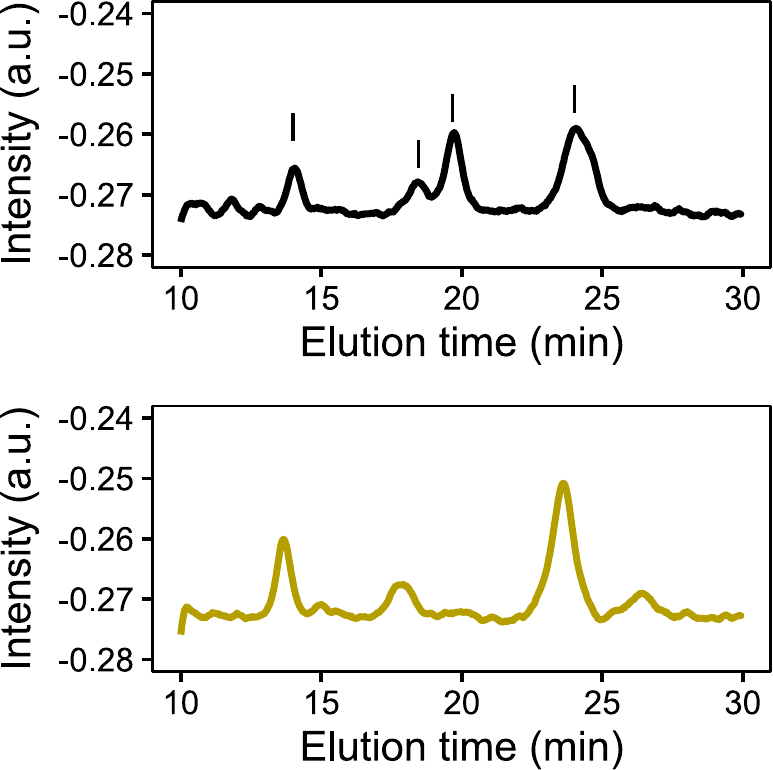}{\SUPPhplccomparison}{%
  Typical HPLC chromatograms of commercial \surfactin{} (top) and
  \surfactin{} produced by our \Bsubtilis{} strain \surfactantstrain{}
  (bottom). The peaks highlighted in the commercial \surfactin{}
  chromatogram were exploited for the construction of the calibration
  plot.}

\SIfigure[width=0.8\columnwidth]{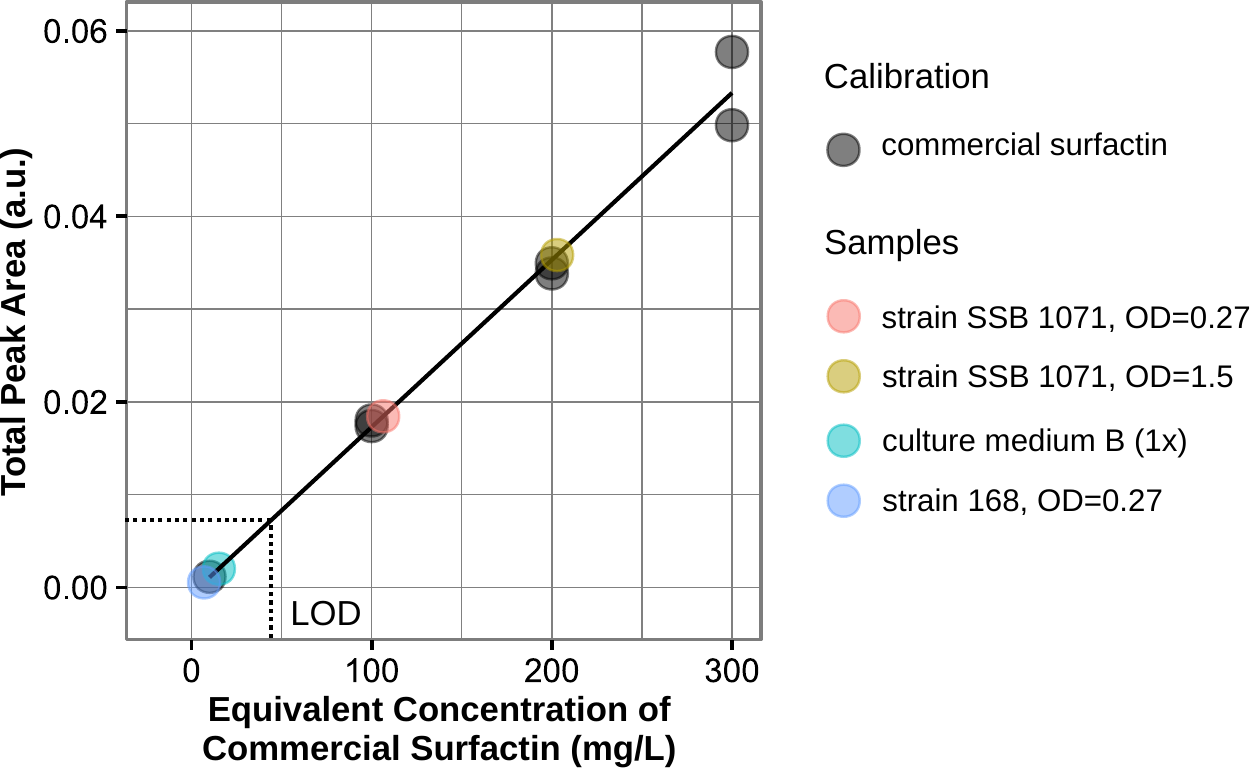}{\SUPPhplccalibration}{%
  Total \surfactin{}-correlated peak intensity vs. commercial
  \surfactin{} concentration calibration plot. LOD indicates the Limit
  Of Detection (see Materials and Methods).}

\SItable{\SUPPhplcconcentrations}{%
  \begin{tabular}{ll}
    \textbf{Sample} & \textbf{Equivalent commercial concentration} \\
    & \textbf{(90\% confidence limits) (mg/L)} \\\hline
    strain \surfactantstrain, OD=0.27 & \num{106+-32} \\
    strain \surfactantstrain, OD=1.5 & \num{202+-32} \\
    B(1x), culture medium & \SI{<44}{ppm} (below LOD) \\
    strain 168, OD=0.27 & \SI{<44}{ppm} (below LOD) \\
  \end{tabular}%
}{%
  Surfactin concentration estimation from HPLC measurements. LOD
  indicates the Limit Of Detection (see Materials and Methods).}

\SIfigure[width=0.8\columnwidth]{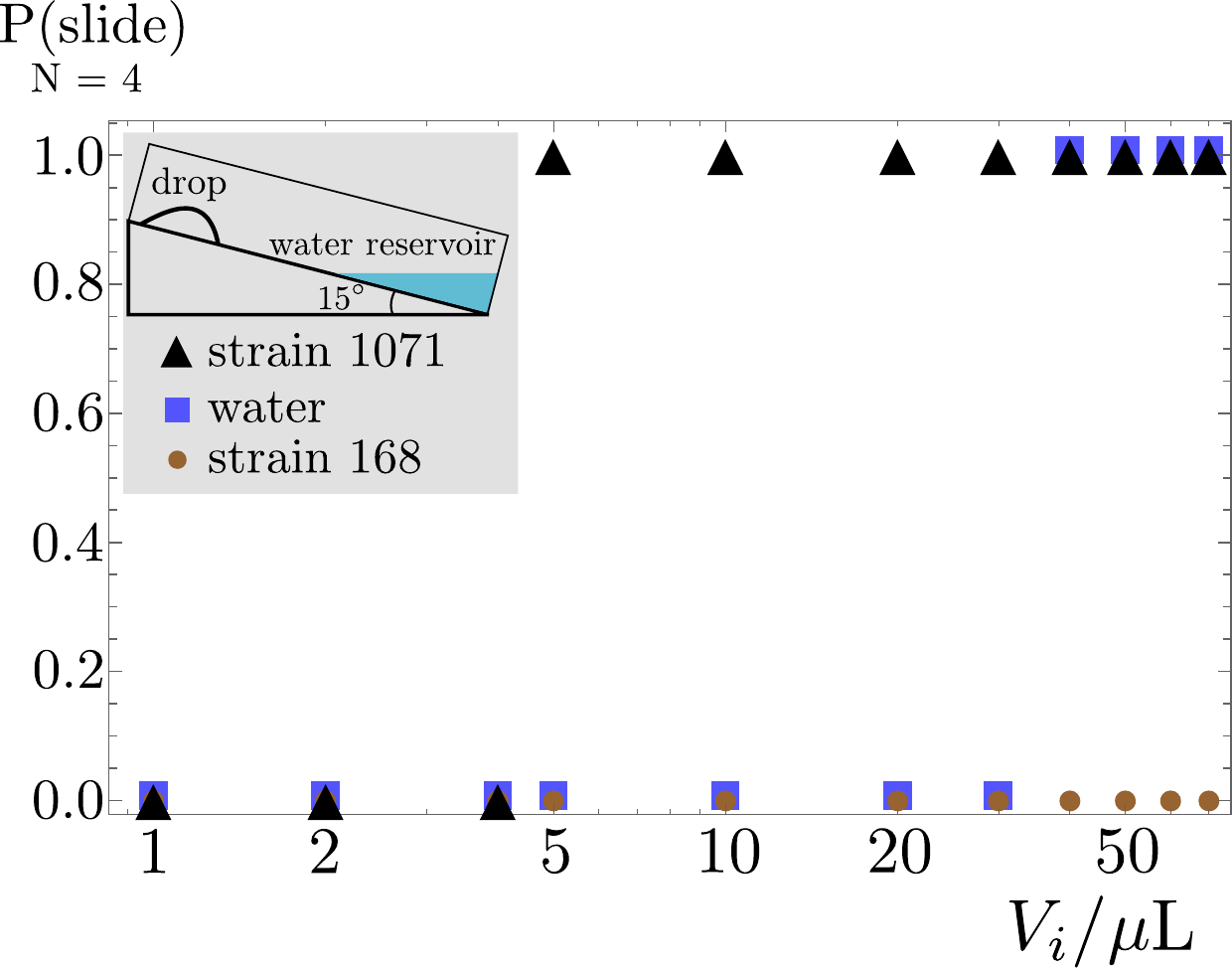}{\SUPPpetridishsliding}{%
  Probability to observe colony surfing in a bare polystyrene Petri
  dish at \ang{15} tilt, closed, and with a water reservoir at the
  bottom, as a function of the initial deposited volume. The surfactin
  producing strain \surfactantstrain{} starts to slide at much smaller
  volumes ($V_i^{\text{crit}} = \SI{4}{\micro\litre}$) when compared
  to water and the surfactin-deficient 168 strain. The experimental
  protocol closely follows that of the experiments on gel. The
  bacterial suspension is prepared using the same protocol (see
  Materials and Methods). Drops of different volume at an initial
  optical density of $\OD=0.27$ are deposited in an inclined clean
  polystyrene Petri dish placed inside a climate chamber
  (T=\SI{30}{\celsius}, RH=\SI{70}{\percent}). \SI{10}{ml} of
  distilled water is added at the bottom of the dish as a source of
  humidity (see inset). Finally the Petri dish cover lid is put in
  place and the chamber is closed. Four drops were measured for each
  condition.}


\end{document}